\documentclass[sigconf]{aamas} 
\usepackage{balance} % for balancing columns on the final page
\usepackage{verbatim}
\usepackage{cuted}
\usepackage{mathtools, bm}
\usepackage{amsfonts,latexsym,color}
\usepackage{algorithm}
\usepackage{algorithmic}
\usepackage{comment}
\usepackage{wrapfig}
\usepackage{amsmath}
\usepackage{comment}

\setcopyright{ifaamas}
\acmConference[AAMAS '23]{Proc.\@ of the 22nd International Conference
on Autonomous Agents and Multiagent Systems (AAMAS 2023)}{May 29 -- June 2, 2023}
{London, United Kingdom}{A.~Ricci, W.~Yeoh, N.~Agmon, B.~An (eds.)}
\copyrightyear{2023}
\acmYear{2023}
\acmDOI{}
\acmPrice{}
\acmISBN{}

\acmSubmissionID{443}

\title[AAMAS-2023 Formatting Instructions]{Parameter Sharing with Network Pruning for Scalable Multi-Agent Deep Reinforcement Learning}

\author{Woojun Kim}
\affiliation{
  \institution{KAIST}
  \city{Daejeon}
  \country{Korea}}
\email{woojun.kim@kaist.ac.kr}

\author{Youngchul Sung}
\affiliation{
  \institution{KAIST}
  \city{Daejeon}
  \country{Korea}}
\email{ycsung@kaist.ac.kr}

\begin{abstract}
Handling the problem of scalability is one of the essential issues for multi-agent reinforcement learning (MARL) algorithms to be applied to real-world problems typically involving massively many agents. For this, parameter sharing across multiple agents has  widely been used since it reduces the training time by decreasing the number of parameters and increasing the sample efficiency. However, using the same parameters across agents limits the representational capacity of the joint policy and consequently, the performance can be degraded in multi-agent tasks that require different behaviors for different agents. In this paper, we propose a simple method that adopts  structured pruning  for a deep neural network to increase the representational capacity of the joint policy without introducing additional parameters. We evaluate the proposed method on several benchmark tasks, and numerical results show that the proposed method significantly outperforms other parameter-sharing methods.

\end{abstract}


\newcommand{\BibTeX}{\rm B\kern-.05em{\sc i\kern-.025em b}\kern-.08em\TeX}

\begin{document}

%%% The following commands remove the headers in your paper. For final 
%%% papers, these will be inserted during the pagination process.

\pagestyle{fancy}
\fancyhead{}

%%% The next command prints the information defined in the preamble.

\maketitle 

%%%%%%%%%%%%%%%%%%%%%%%%%%%%%%%%%%%%%%%%%%%%%%%%%%%%%%%%%%%%%%%%%%%%%%%%

\section{Introduction}

Multi-agent reinforcement learning (MARL) is gaining attention as an important research direction  for tackling many real-world decision-making tasks such as traffic management and autonomous driving.  Recently, many  issues  for MARL have vigorously been studied to advance MARL theories and algorithms, including value factorization \cite{rashid2018qmix, sunehag2018value}, communication \cite{foerster2016learning, kim2020communication, kim2019message}, subgoal generation \cite{jeon2022maser}, and correlated exploration \cite{mahajan2019maven, kim2023variational}. 
In order to apply these MARL algorithms to real-world problems with many agents and constrained resources, one needs to  handle the problem of scalability and sample inefficiency in most cases. 
This is because these algorithms typically have a high variance for gradient estimation and require increasing computational cost and memory as the number of agents increases.
One way to address the problem of scalability and sample inefficiency is  \textit{parameter sharing},  commonly adopted in most MARL algorithms \cite{rashid2018qmix, foerster2016learning,foerster2018counterfactual,jeon2022maser}. 
Parameter sharing uses only one parameterized function such as a deep neural network for all agents' policies or critics, and increases sample efficiency yielding fast training. However, for naive parameter sharing using a single common function,  agents choose the same action if their observations are the same and critics estimate the expected return to be the same even if agents receive individual rewards. As a result, naive parameter sharing has a limited representational capacity of the joint policy and critics, and this can degrade performance in heterogeneous multi-agent environments or environments that require different behaviors for different agents.
Thus, to allow  multiple agents even with the same observations to perform different actions, the concatenation of the agent's observation  and a one-hot vector encoding agent indication is widely  used for recent MARL algorithms \cite{rashid2018qmix, mahajan2019maven, jeon2022maser}. 
Although this method improves performance compared with naive parameter sharing, the representational capacity is still limited because the neurons of a deep neural network for function approximation are shared across all agents but only agent-specific perturbation is added to a hidden layer. 
Furthermore, it is difficult for this one-hot-vector-based method to control the amount of parameter sharing among agents, and it even introduces an additional parameter of one hot vector.

In this paper, we propose  a novel structured network pruning method for parameter sharing (SNP-PS) in multi-agent deep reinforcement learning so as to increase the representational capacity of neural networks parameterizing actors and critics without introducing additional parameters while keeping high sample efficiency.  Inspired by the lottery ticket hypothesis \cite{su2020sanity}, we conjecture  a new hypothesis named the lottery group ticket hypothesis, which assumes  the existence of a group of pruned subnetworks that allow multiple agents to be identifiable and improve performance compared with the naive parameter sharing. Then, we  provide a neural network pruning method to obtain such a group of subnetworks called winning group tickets. The obtained group of subnetworks is  used for the policies or critics of multiple agents. Since the policies or critics of the proposed method shares only subsets of  the parameters, they can yield diverse joint actions. 
The proposed method is very easy to implement and can be combined with other parameter-sharing methods. 
Furthermore, the proposed method can control the number of shared parameters by adjusting the pruning ratio without introducing additional parameters. 
We evaluated the proposed parameter-sharing method on top of QMIX and multi-agent advantage actor-critic in several multi-agent environments. Numerical results show that the proposed method noticeably enhances  both training speed and the final performance compared with  other parameter-sharing methods.

% neuron units are shared across entire agents or some of the agents,  or only one agent.

%%%%%%%%%%%%%%%%%%%%%%%%%%%%%%%%%%%%%%%%%%%%%%%%%%%%%%%%%%%%%%%%%%%%%%%%
\section{Background}

\subsection{Environment Model}

\textbf{A Partially Observable Markov Game} A  general multi-agent task is typically modeled as a Partially Observable Markov Game (POMG), also known as a partially observable stochastic game. A POMG is represented by the tuple $<\mathcal{N}, \mathcal{S},\{\mathcal{A}_i\}_{i=1}^{N},\{\Omega_i\}_{i=1}^{N}, \mathcal{T},\mathcal{O},$ $\{\mathcal{R}_i\}_{i=1}^{N}>$, where $\mathcal{N}$ is the set of agents, $\mathcal{S}$ is the state space of the environment, $\boldsymbol{\mathcal{A}}=\prod_{i=1}^{N} \mathcal{A}^i$ is  the joint action space, and $\boldsymbol{\Omega}=\prod_{i=1}^{N} \Omega^i$ is the joint observation  space. At each time step $t$, Agent $i$ perceives local observation $o_t^i \in \Omega_i$, which is determined from the global state $s_t \in \mathcal{S}$ according to the observation probability $\mathcal{O} : \mathcal{S} \times \boldsymbol{\mathcal{A}} \times \boldsymbol{\Omega} \rightarrow [0,1]$,  and determines and executes action $a_t^i$ based on its local history. The joint action $\boldsymbol{a_t}=(a_t^1,\cdots, a_t^N)$ yields next state $s_{t+1}$ according to  the transition probability $\mathcal{T}:\mathcal{S}\times \boldsymbol{\mathcal{A}} \times \mathcal{S}\rightarrow [0,1]$. Then,  Agent $i$ receives local reward $r_t^i$ according to  its own reward function $\mathcal{R}_i : \mathcal{S}\times \boldsymbol{\mathcal{A}} \rightarrow {\mathbb{R}}$ and  next observation $o_{t+1}^i$. This procedure is repeated and each agent aims to find its own policy that maximizes the expected return defined as $R_t^i = \sum_{t'=t}^{\infty} \gamma^{t'} r_{t'}^i$ where $\gamma \in [0,1]$ is the discounting factor. Here, Agent $i$'s policy is conditioned on its own observation-action history $\tau^i\in (\Omega_i \times \mathcal{A}_i)^*$. The expected return of one agent depends on other agents' policies, and consequently, the notion of equilibrium, e.g. \textit{Nash equilibrium} naturally arises in the objective function of all agents. %In addition, depending on the reward function, the task is divided into three categories: cooperative, competitive, and mixed cooperative-competitive environments.

If all agents receive the same reward, i.e. team reward, the task is considered as a fully cooperative setting and the model reduces to a decentralized partially observable Markov decision process (Dec-POMDP). In Dec-POMDP, all agents optimize their policies to maximize the common objective function $J(\boldsymbol{\pi})=\mathbb{E}_{\pi}(\sum_{t=0}^{\infty} \gamma^t r_t)$, where $\boldsymbol{\pi}=(\pi^1,\cdots, \pi^N)$ is the joint policy.

%Since the joint action yields the team reward, it is difficult to pinpoint the contribution of each agent to the team reward  we cannot assign the proper reward to encourage the agent to maximize the global reward, which is called a multi-agent credit assignment problem. 

\subsection{Multi-Agent Reinforcement Learning}

One simple approach to MARL is using a centralized controller which basically represents the joint policy. This approach requires communication channels among agents to handle  partial observability and suffers from the curse of dimensionality, i.e., the volume of the input space grows exponentially as  the number of agents increases. 
An alternative approach is independent learning in which each agent has a decentralized policy and learns the policy independently while treating other agents as a part of the environment. 
The representative example is independent Q-learning (IQL), which is an extension of Q-learning to multi-agent settings \cite{tan1993multi}. However, IQL limits the learning of coordinated behaviors due to the negligence of the influence of other agents, and thus the centralized action-value function conditioned on the environmental state and the joint action has been adopted \cite{lowe2017multi, foerster2018counterfactual, rashid2018qmix}. 
%In IQL, the action-value function of each agent is only conditioned on local information, and consequently, it can avoid the curse of dimensionality but limits the learning of coordinated behaviors due to the negligence of the influence of other agents.
%Thus, in order to capture the influence of other agents, the centralized action-value function conditioned on the environmental state and the joint action has been adopted \cite{lowe2017multi, foerster2018counterfactual, rashid2018qmix}. 
A well-known example is QMIX, which  decomposes the joint action-value function $Q_{JT}(s,\boldsymbol{\tau}, \boldsymbol{a})$ into a non-linear combination of individual action-value functions with a monotonic constraint as follows:
\begin{align}
    Q_{JT}(s,\boldsymbol{\tau}, \boldsymbol{a}) = f_{mixing}(s, &Q^1(\tau^1, a^1), \cdots, Q^N(\tau^N, a^N)), \nonumber \\ &  \mbox{s.t.}~\frac{\partial Q_{JT}(s,\boldsymbol{\tau}, \boldsymbol{a})}{\partial Q^i(\tau^i, a^i)} \geq 0, ~ \forall i \in \mathcal{N},
\end{align}
where $Q_i(\tau^i, a^i)$ is the individual action-value function of Agent $i$ and $f_{mixing}$ is the mixing network which is trained to satisfy the monotonic constraint by imposing non-negativity on  its weights.   The individual action-value functions are parameterized by multi-layer perceptrons (MLPs) and a gated recurrent unit (GRU). For practical implementation and improved sample efficiency, the parameters are shared across all agents and an encoded one-hot vector is added to the input for agent indication. The joint action-value function is parameterized by $\theta$  and  trained to minimize the temporal difference error with the loss function, given by
\begin{align}
    \mathcal{L}(\theta) = \mathbb{E}\Big[ (r+ \gamma \max_{\boldsymbol{a'}} Q_{JT}(s',\boldsymbol{\tau}', \boldsymbol{a}';\theta^{-}) - Q_{JT}(s,\boldsymbol{\tau}, \boldsymbol{a};\theta))^2  \Big]
\end{align}
where $s', \boldsymbol{\tau}'$ and $\theta^{-}$ are the next state, the next joint history $\boldsymbol{\tau}'=(\tau^1, \cdots, \tau^N)$, and the parameter of the target network, respectively.  In this paper, we consider QMIX as the baseline algorithm to test the proposed parameter-sharing method.  In addition to QMIX, which is a value-based approach, we also consider an actor-critic-based multi-agent algorithm which extends advantage actor-critic (A2C) to multi-agent setting considered in \cite{christianos2021scaling}.

\begin{figure*}[t]
\begin{center}
\begin{tabular}{ccc}
     % uncomment the next lines, and give the right ps files
     %\includegraphics[width=0.23\textwidth]{figures/MW_N2.png} &
     \includegraphics[width=0.23\textwidth]{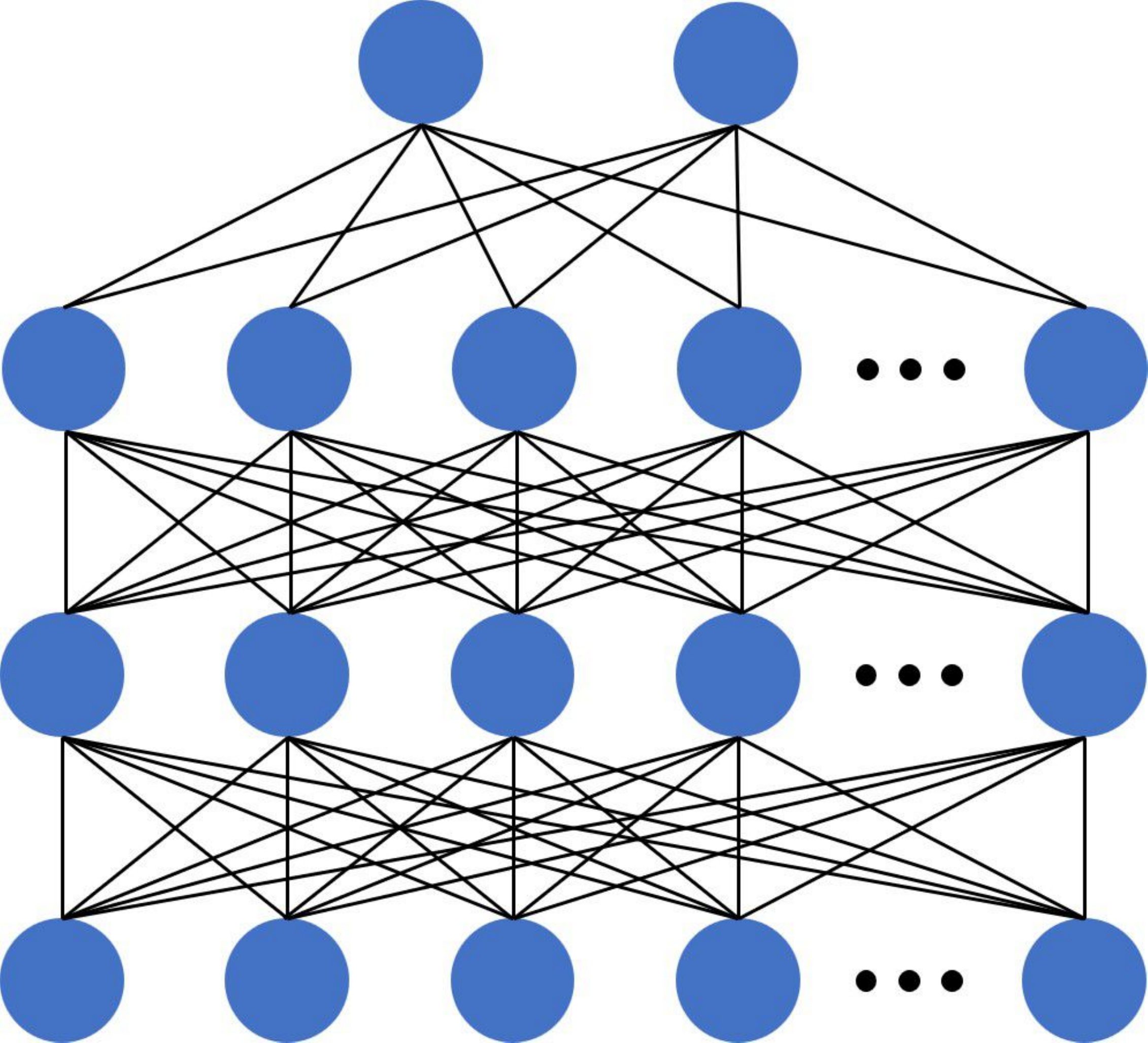} \hspace{10ex} &
     \includegraphics[width=0.23\textwidth]{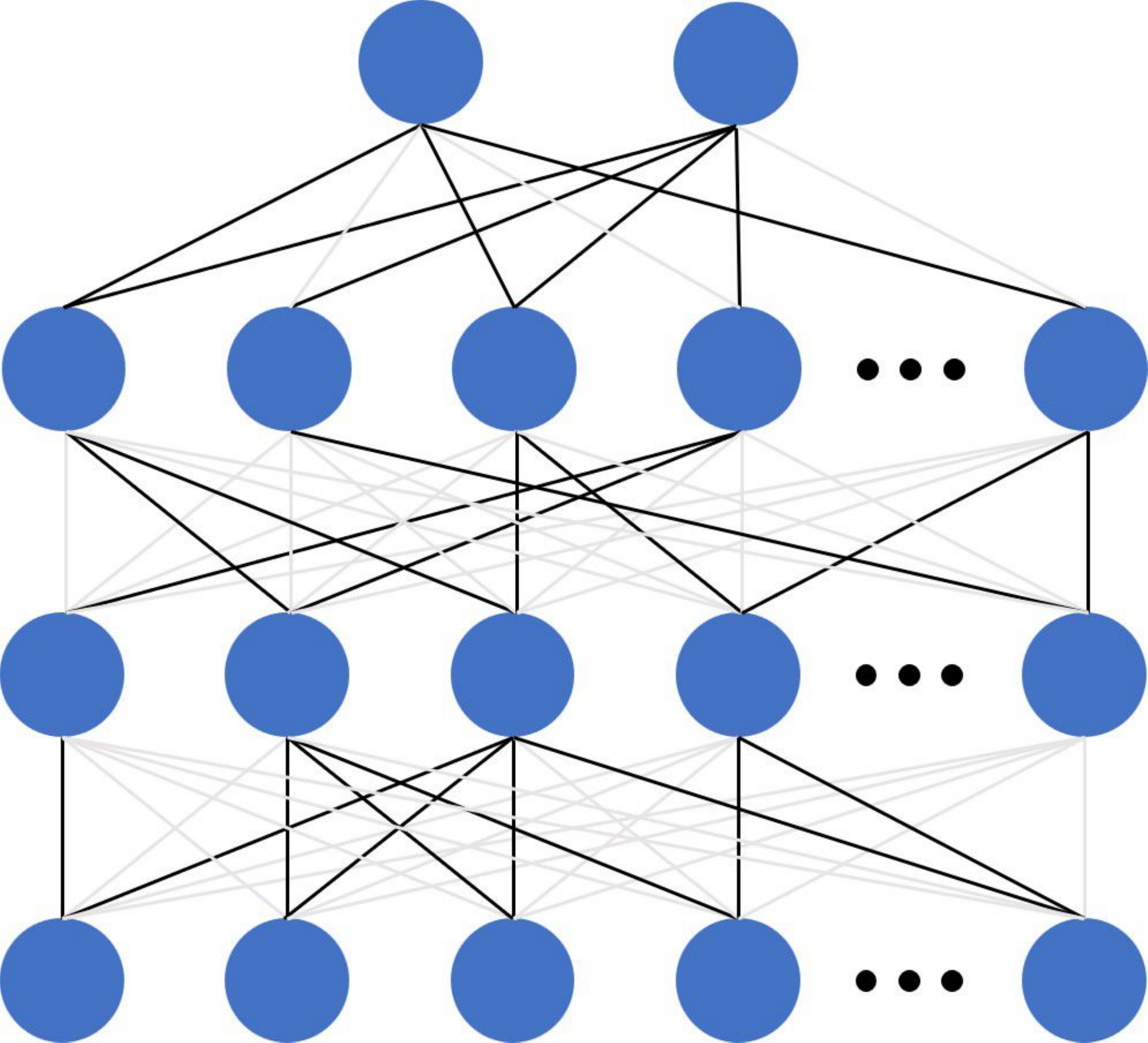} \hspace{10ex} &
     \includegraphics[width=0.23\textwidth]{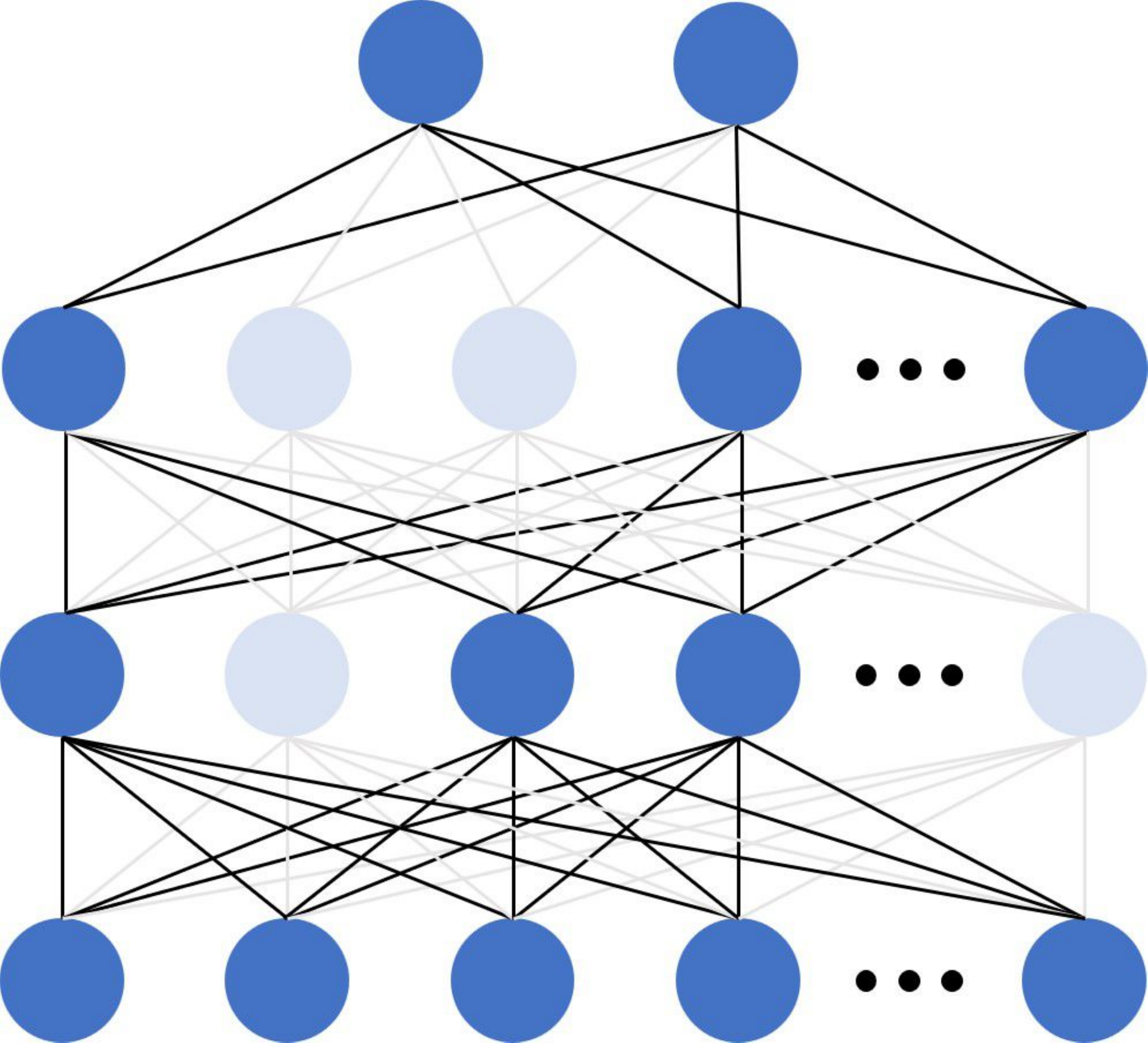} \\
     (a) \hspace{10ex} & (b) \hspace{10ex} & (c)
\end{tabular}
 \caption{(a) An original dense neural network, (b) a weight-pruned network, and (c) a neuron-pruned network}
\label{fig:neuralnetworks}
\end{center}
\end{figure*}
%%%%%%%%%%%%%%%%%%%%%%%%%%%%%%%%%%%%%%%%%%%%%%%%%%%%%%%%%%%%%%%%%%%%%

\subsection{Neural Network Pruning}

Neural network pruning removes parameters such as weights and/or neurons to build a smaller neural network without performance degradation. Recently, neural network pruning has widely been studied for its applicability to resource-constrained tasks. The typical process of neural network pruning consists of three steps: 1) training a dense network, 2) pruning the network, and 3) fine-tuning the pruned network. By iterating this procedure, we can obtain a pruned network, which is often a sparse neural network. This method is called \textit{pruning after training} (PaT) \cite{wang2021recent}. 
In order to make the procedure simpler and reduce the computational cost, pruning at initialization (PaI) is gaining increasing attention under the lottery ticket hypothesis (LTH), which assumes that a randomly initialized dense network contains a subnetwork (called a winning ticket) that can achieve comparable performance to that of the original dense network \cite{frankle2018lottery}. Under the hypothesis, to find a  winning ticket, \citet{su2020sanity} proposed using a random ticket that prunes the weights of a deep neural network randomly while keeping the ratio of pruned weights per layer. Surprisingly, it is shown that the performance of such a randomly pruned network on VGG-19 on CIFAR-10 degrades only $1\%$ when the pruning ratio of the randomly initialized network is $98\%$, i.e., the weights are removed by $98\%$ \cite{su2020sanity}. Although  the lottery ticket hypothesis has not been proved yet due to its theoretical difficulty, practical applications of random pruning indeed yield a subnetwork with comparable performance and hint at the possible validity of the hypothesis.

Neural network pruning methods can be divided into two categories: unstructured pruning and structured pruning. Unstructured pruning removes weights individually by fixing them to zero. On the other hand, structured pruning removes a group of weights together. For example, structured pruning  removes all weights that are connected to a neuron or to filters or channels in a convolutional neural network. Fig. \ref{fig:neuralnetworks} illustrates network pruning. Fig. \ref{fig:neuralnetworks}(a), (b) and (c) show an original dense network, weight-pruned network, and neuron-pruned network, respectively. 
A Weight-pruned network and a neurons-pruned network are examples of unstructured pruning and structured pruning, respectively.

\textbf{Notations} We denote  a deep neural network as a parameterized function $f(x; \theta)$, where $x$ is the input and  $\theta$ is the trainable parameters including weights and biases in the neural network. 
Then, for an original dense deep neural network $f(x;\theta)$, network pruning produces a new neural network model $f(x; \theta \odot M)$, where $M=\{0,1\}^{|\theta|}$ is a binary mask for pruning and $\odot$ denotes the Hadamard product, i.e.,  elementwise product.

%%%%%%%%%%%%%%%%%%%%%%%%%%%%%%%%%%%%%%%%%%%%%%%%%%%%%%%%%%%%%%%%%%%%%%%%

\section{Related Works}

\subsection{Scalability in Multi-Agent Reinforcement Learning}

MARL algorithms often encounter the curse of dimensionality, which arises from the increasing input space of the parameterized function as the number of agents increases. One example is the use of centralized critic, which is conditioned on the joint action and the environment state or the joint observation. It has widely been  considered to address the non-stationarity problem of MARL \cite{lowe2017multi, foerster2018counterfactual}. Here, the volume of the joint action and observation spaces grows exponentially, and consequently, it is difficult to train a deep neural network approximating the centralized critic with a large number of agents \cite{kim2019message}. To address this problem, \citet{iqbal2019actor} proposed  an attention-based centralized critic which can be scalable to the number of agents. \citet{yang2018mean} proposed mean-field actor-critic which uses the mean-field theory to estimate a centralized critic based on local critics capturing  interaction. The curse of dimensionality also arises when  inter-agent communication is allowed. In this setting, each agent generates and sends a message and learns coordinated action based on received messages \cite{kim2020communication, foerster2016learning}. 
Then, the dimension of  received messages increases as the number of agents increases. \citet{kim2019message} proposed an efficient training method to handle the increasing dimension of  messages and it can also handle the aforementioned problem of the centralized critic. 

\begin{figure}[h]
\begin{center}
\begin{tabular}{c}
     % uncomment the next lines, and give the right ps files
     %\includegraphics[width=0.23\textwidth]{figures/MW_N2.png} &
     \includegraphics[width=0.33\textwidth]{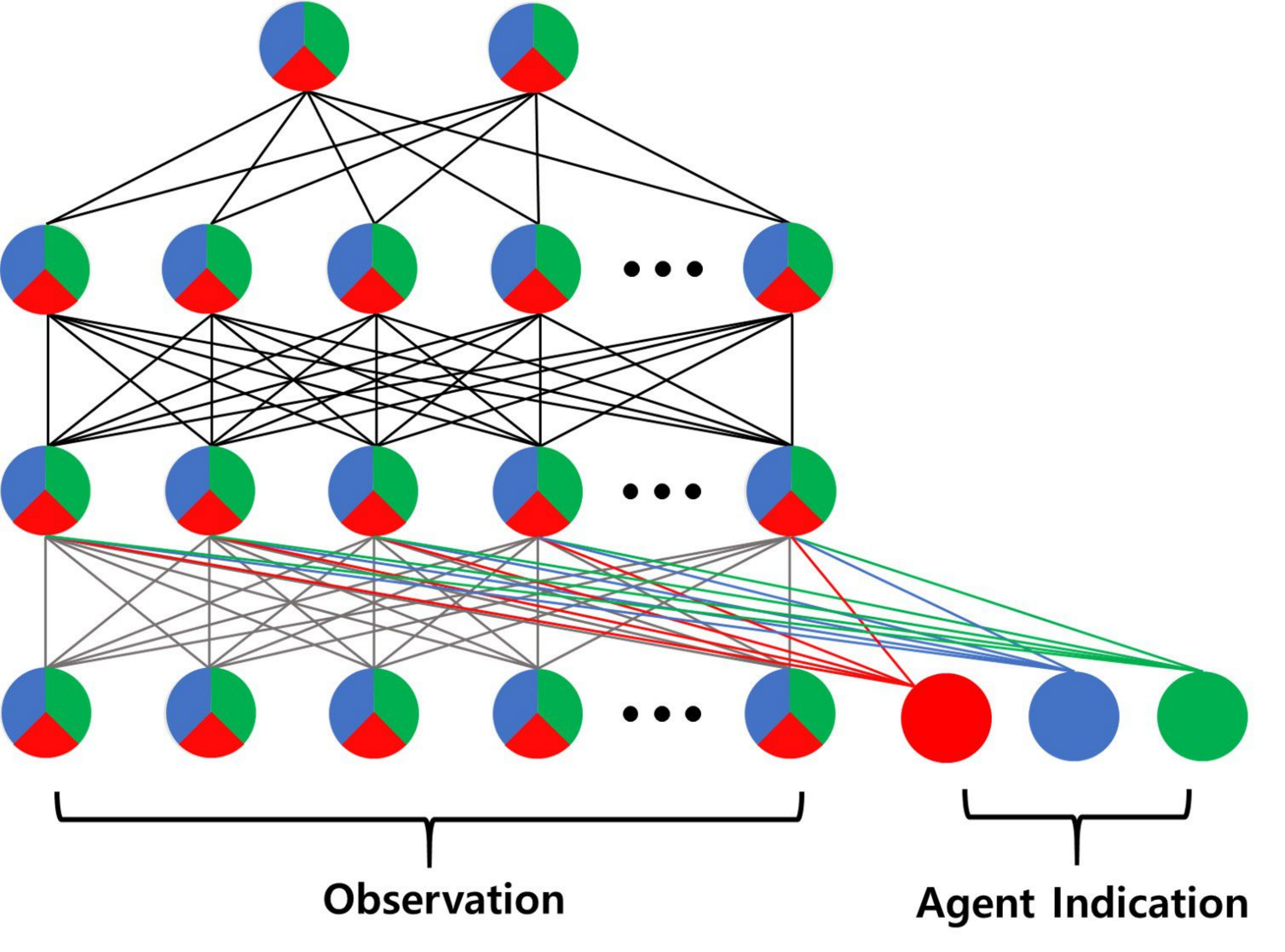} \\
     %(a) \vspace{2ex} \\ 
     %\includegraphics[width=0.14\textwidth]{figures/onehot1.pdf} \hspace{1ex}
     %\includegraphics[width=0.14\textwidth]{figures/onehot2.pdf} \hspace{1ex}
     %\includegraphics[width=0.14\textwidth]{figures/onehot3.pdf}\\
     %\hspace{-4ex} (b) \hspace{17ex} (c) \hspace{17ex} (d) 
\end{tabular}
\vspace{-1ex}
 \caption{Neural network architecture for parameter sharing with one-hot vector (the same network is used for all agents except the one-hot vector value change):  Note that a different one-hot vector only affects the input to the first hidden-layer through the link weights from the activated one-hot vector unit to the first hidden layer, and the network from the observation input to the output layer is the same for all agents.}
\label{fig:onehotnetwork}
\end{center}
\vspace{-4ex}
\end{figure}

In addition to the curse of dimensionality, the increase in  the number of trainable parameters with respect to (w.r.t.) the number of agents is also an issue. For an $N$-agent actor-critic algorithm with deep neural network-based function approximation, we need $2N$ deep neural networks to parameterize all policies and critics. Thus, the required memory size grows linearly w.r.t. the number of agents.   Furthermore, the separate approximation of all agents' networks limits the sample efficiency of learning, because in the separate approximation case, each agent's critic and policy can only use its own observation  for training.   To address the aforementioned increasing hardware and sample inefficiency problem, \textit{parameter sharing} has widely been used in MARL research \cite{rashid2018qmix, kim2023variational}. 
By sharing the parameters across agents, the required memory size does not grow w.r.t. the number of agents and the training time can be reduced. 
However, parameter sharing limits the representational capacity of the joint policy and critic, and consequently, agents with the same  observation-action histories  will act in the same way. This can cause performance degradation in multi-tasks that require diverse behaviors, e.g., different behaviors for different agents. 
To avoid this limitation of simple parameter sharing,  most recent MARL algorithms combine parameter sharing with agent indication to the observation \cite{rashid2018qmix, kim2023variational}. 
For agent indication, one-hot vector is commonly used, as shown in Fig. \ref{fig:onehotnetwork}.  It was shown that due to the impact of the one-hot vector on the first hidden layer, parameter sharing using one-hot vector yields better performance than simple parameter sharing. \citet{terry2020parameter} proved that agent indication allows agents' policies to converge to  optimal ones.

\subsection{Network Pruning in Reinforcement Learning}

\citet{graesser2022state} investigated the application of existing network pruning techniques to reinforcement learning. They found that sparse training based on network pruning often performs better than the dense network with the same number of parameters and increases the robustness against  observation noise. \citet{livne2020pops} proposed transfer-learning-based network pruning to prevent the pruned network from performance degradation. They first train a teacher network, i.e., a dense neural network, and then prune and regenerate the weights. \citet{sokar2021dynamic} proposed a dynamic sparse training method to reduce the required resources, e.g. memory and computational cost, while improving the performance. 
However, these works are proposed for single-agent reinforcement learning and are based on unstructured network pruning  to reduce the trainable parameters with keeping performance degradation.

%%%%%%%%%%%%%%%%%%%%%%%%%%%%%%%%%%%%%%%%%%%%%%%%%%%%%%%%%%%%%%%%%%%%%
\begin{figure*}[t]
\begin{center}
\begin{tabular}{c}
     % uncomment the next lines, and give the right ps files
     %\includegraphics[width=0.23\textwidth]{figures/MW_N2.png} &
     \includegraphics[width=0.975\textwidth]{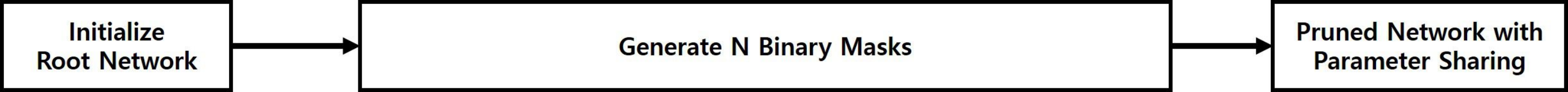} \vspace{1ex} \\
     \hspace{-2ex}
     \includegraphics[width=0.15\textwidth]{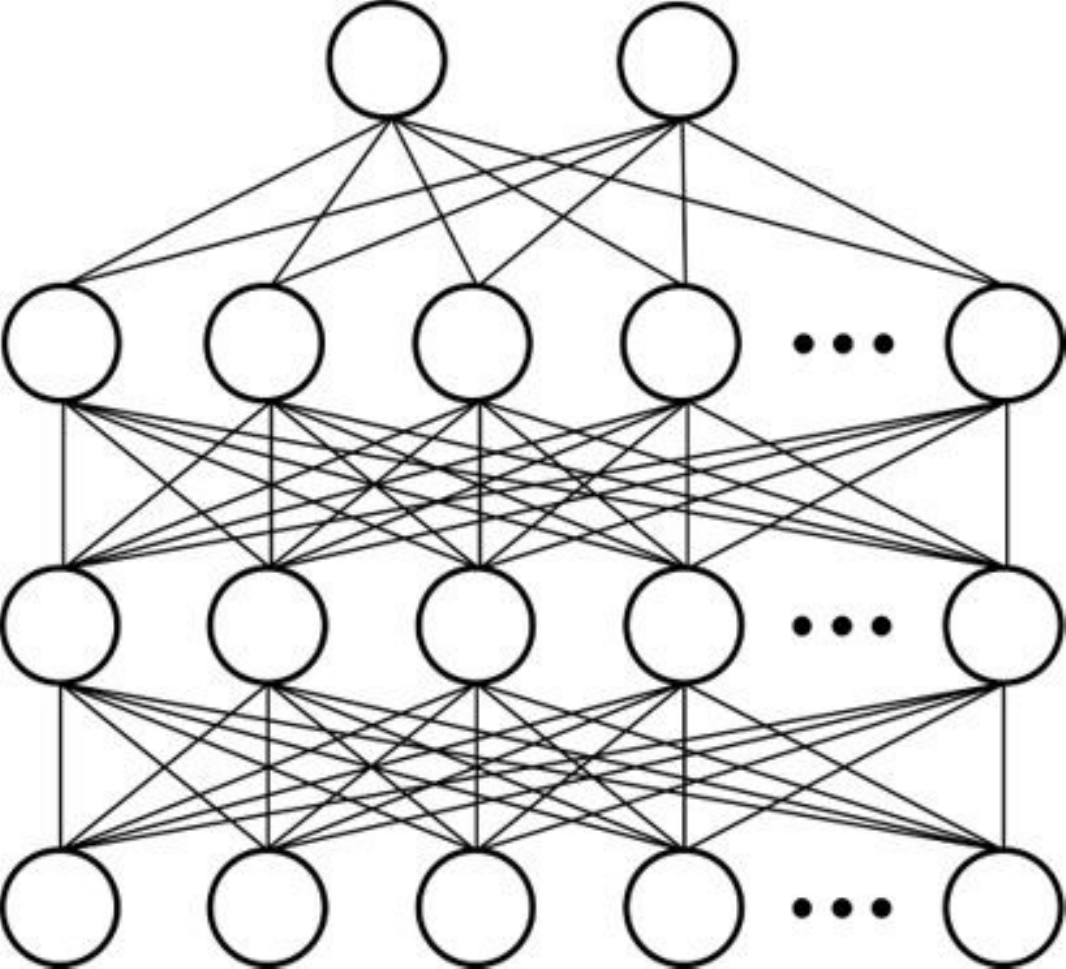} \hspace{14ex} ~~
     \includegraphics[width=0.45\textwidth]{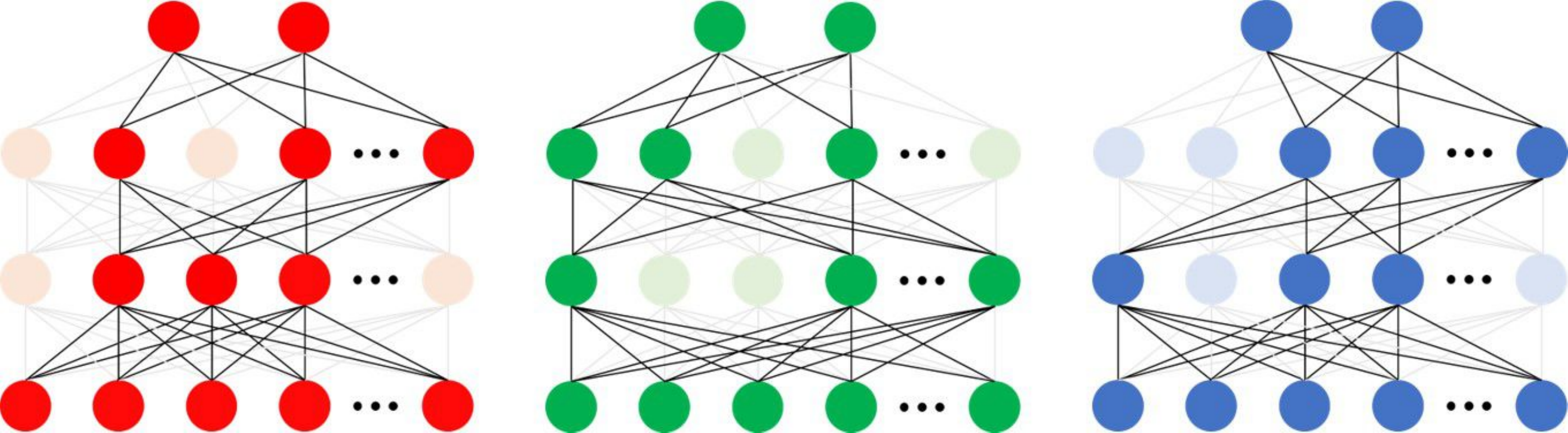} \hspace{12ex} ~~
     \includegraphics[width=0.15\textwidth]{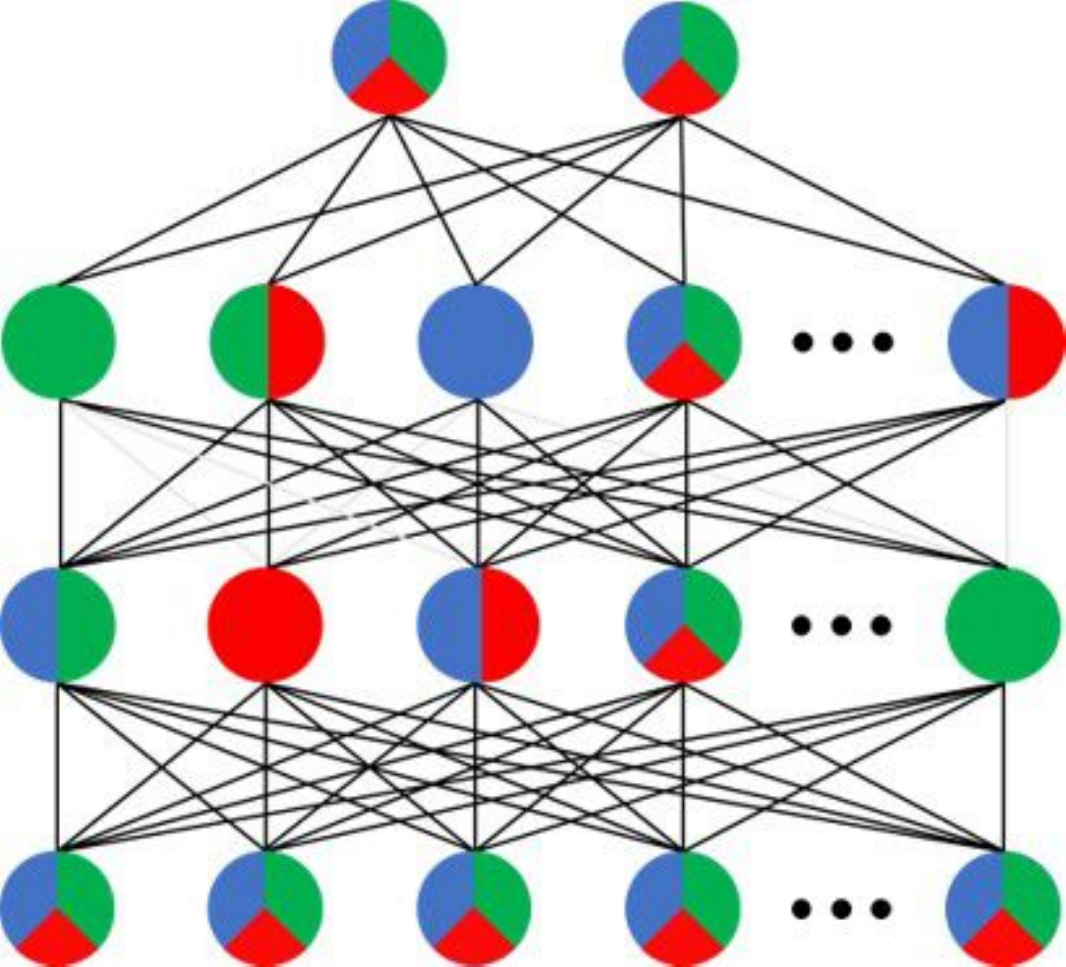} \vspace{1ex} \\
     (a) \hspace{47ex} (b) \hspace{49ex}  (c)
\end{tabular}
 \caption{The procedure of generating $N=3$ subnetworks: (a) randomly-initialized dense neural network. (b) $N=3$ subnetworks are generated from the same dense network independently. Each colored neural network represents different agents' subnetworks. (c) the generated subnetworks share some subsets of parameters across multiple agents. Each neuron is colored with the colors of agents who share the corresponding neuron.}
\label{fig:procedure}
\end{center}
\end{figure*}

%%%%%%%%%%%%%%%%%%%%%%%%%%%%%%%%%%%%%%%%%%%%%%%%%%%%%%%%%%%%%%%%%%%%%%%%

\section{Methodology}

We consider neural network pruning  to reinforce agent indication for learning diverse actions in multi-agent reinforcement learning.  First, we present our motivation. Then,   inspired by \cite{su2020sanity}, we provide our conjecture named the lottery group ticket hypothesis relevant to  multi-agent deep reinforcement learning network pruning,  and propose a structured network pruning-based parameter sharing method to identify winning group tickets of the conjecture.

\subsection{Motivation}

To increase the representational capacity of simple  parameter sharing, the concatenation of  agents' observation and an agent-specific one-hot vector, called one-hot encoding agent indication, has been used due to its simplicity and performance improvement. 
However, the one-hot encoding agent indication still has  limited representational capacity for different agents. Fig. \ref{fig:onehotnetwork} shows the neural network architecture of one-hot encoding agent indication. The input layer consists of one-hot vector of size $N$ and observation input.  
As seen in the figure, the weight and bias parameters are the same for all agents except the weights from the one-hot vector input unit to the first hidden layer.  For Agent $i$, $o^i$ is applied to the observation input, and the $i$th element of the one-hot vector input is activated to one and all other elements of  the one-hot vector input  are deactivated as zero.  Thus, the first hidden-layer feature vector is the sum of the observation feature and the agent indication feature.   One can think the first hidden-layer vector is the observation feature added by an agent-specific perturbation.  Different agent representational capacity comes only from this perturbation vector. 
No neurons are agent-specific and the common network is shared by all agents  except the one-hot input and corresponding links.   
Another problem of the one-hot encoding agent indication is that it is difficult to explicitly control the amount of parameter sharing due to its fixed architecture. Depending on tasks, one may want similar or quite disparate actions by agents for similar observations. For this, it is necessary to control  the amount of parameter sharing. Furthermore, the input size of the one-hot encoding agent indication method linearly increases as the number of agents increases. This may be problematic for MARL with a huge number of agents.

Instead of adding one-hot vector, \citet{christianos2021scaling} proposed a parameter-sharing method named selective parameter sharing (SePS). SePS divides all agents into subgroups that share parameters based on a clustering algorithm before training. In SePS, unlike one-hot encoding agent indication, the agents only in the same cluster share parameters, and this can increase representational capacity. For this, they encode agent indication based on a variational autoencoder (VAE) and then  use the encoded hidden vector of VAE for the clustering algorithm.  However, SePS requires  networks as many as the number of clusters, and this  decreases sample efficiency and increases the required memory resources. In addition, the method  depends on the performance of clustering since the clustered agents are fixed during training.

Circumventing the limitations of the previous methods, we here aim to devise  a parameter-sharing method that  increases the representational capacity and the sample efficiency without introducing additional parameters. For this, we exploit  structured neural network pruning. We start with our conjecture named the lottery group ticket hypothesis.

\subsection{The Lottery Group Ticket Hypothesis} 

%We propose a structured network pruning-based parameter sharing to make agents parameterized by the same parameters identifiable. Inspired by the lottery ticket hypothesis (LTH), we state the following the lottery correlated tickets hypothesis.
Inspired by the lottery ticket hypothesis (LTH) \cite{su2020sanity}, we conjecture  the following hypothesis. 

\textbf{The Lottery Group Ticket Hypothesis (LGTH):} ~~ \textit{A randomly-initialized dense network with sufficient size contains a group of subnetworks that allow multiple agents to be identifiable and yields the return performance comparable to  the full dense network with simple parameter sharing if subnetworks in place of the full dense network are used for policies or critics in multi-agent reinforcement learning.}

Formally speaking, the hypothesis can be stated as follows. Suppose that MARL agents parameterize their policies based on the common function $f(x;\theta)$ and their own masks $\{M_i\}_{i=1}^{N}$. We obtain Agent $i$'s policy as  $f(x;\theta\odot M_i)$, where $\odot$ is  element-wise product.  Then, the lottery group ticket hypothesis conjectures that for a sufficient-size network and up to a certain $N$,  there exist masks $\{M_i\}_{i=1}^{N}$ such that  $\{f(x;\theta\odot M_i)\}_{i=1}^{N}$ are different for the same input $x$ (identifiability) and $J(\prod_{i=1}^{N} f(x;\theta_{final}))\leq J(\prod_{i=1}^{N} f(x;\theta_{final}\odot M_i))$ (comparable performance to the full dense network), where $\theta_{final}$ is the parameters after training finishes and $J(\cdot)$ is the objective function of MARL. We refer to a group of subnetworks $\{M_i\}_{i=1}^{N}$ that satisfy the LGTH as winning group tickets.

Like LTH \cite{su2020sanity}, we do not have proof for LGTH, and both LTH and LGTH are mere conjectures at this point. However, our parameter sharing method developed hereafter under the conjecture indeed yields a group of subnetworks satisfying both identifiability and comparable performance conditions, indicating the possible validity of the conjecture, and proof of LTH and LGTH remains as a nontrivial future work.

\begin{comment}
\begin{algorithm}[t]
   \caption{Generate Subnetworks }
   \label{alg:pnps}
\begin{algorithmic}
   \STATE Initialize parameters of root networks for actor and critic $\theta_{\mu}, \theta_Q,$
   \FOR{$i=1,2,\cdots, N$}
   \FOR{$l=1,2,\cdots, H$}
   \STATE Select the fixed ratio of neurons to be pruned
   \STATE Generate the binary mask $M_i^l$
   \ENDFOR
   \STATE Set the actor and critic network using the binary mask, $f(x;\theta \odot M_i$), where $M_i=(M_i^1,\cdots, M_i^l)$
   \ENDFOR
\\
\end{algorithmic}
\end{algorithm}
\end{comment}

\begin{comment}
\begin{algorithm}[t]
   \caption{Generate Winning Group Tickets}
   \label{alg:pnps}
\begin{algorithmic}
   \STATE Initialize parameters of dense networks for actor and critic $f_\pi(x;\theta_{\pi}), f_Q(x;\theta_Q)$
   \FOR{$i=1,2,\cdots, N$}
   \FOR{$l=1,2,\cdots, L$}
   \STATE Select the fixed number of neurons to be pruned and generate the corresponding binary masks $M_{i}^{\pi,l}$ for policy
   \STATE Select the fixed number of neurons to be pruned and generate the corresponding binary masks $M_{i}^{Q,l}$ for critic
   \ENDFOR
   \STATE Set the actor and critic network for Agent $i$ using the binary masks, $f_{\pi}(x;\theta_{\pi} \odot M^{\pi}_i$) and $f_Q(x;\theta_Q  \odot M^Q_i)$, where $M^{\pi}_i=(M_i^{\pi,1},\cdots, M_i^{\pi, L})$ and $M^{Q}_i=(M_i^{Q,1},\cdots, M_i^{Q, L})$
   \ENDFOR
\\
\end{algorithmic}
\end{algorithm}
\end{comment}

\begin{algorithm}[t]
   \caption{Generate Winning Group Tickets}
   \label{alg:snpps}
\begin{algorithmic}
   \STATE Initialize parameters of root networks for actor and critic
   \FOR{$Agent=1,2,\cdots$}
   \FOR{$Layer=1,2,\cdots$}
   \STATE Select the fixed number of neurons to be pruned randomly
   \STATE Generate the corresponding binary masks for the actor and critic
   \ENDFOR
   \STATE Initialize neural networks for the policy and critic based on the binary masks 
   \ENDFOR
\\
\end{algorithmic}
\end{algorithm}

\subsection{Structrued Network Pruning with Parameter Sharing}

We now propose our pruning-at-initialization (PaI)-based structured network pruning method for parameter sharing (SNP-PS) to construct a collection of winning group tickets for MARL.  Before training, we first build a randomly-initialized dense neural network and then randomly prune a fixed number of neurons per layer to generate a subnetwork. We repeat this pruning $N$ times independently to generate $N$ subnetworks from the same dense network, as seen in Figs. \ref{fig:procedure} (a) and (b). 
Random network pruning at initialization with a fixed pruning ratio was  successfully used to obtain  a winning ticket for a classifier under LTH \cite{su2020sanity}.  
Whereas
 \citet{su2020sanity} adopted  unstructured pruning, we use structured pruning suitable to obtain multiple policies or critic subnetworks for MARL. In our pruning method, we 
 remove the group of weights that are connected to the same neuron altogether. 
Furthermore,  we generate $N$-subnetworks so that they share some subsets of parameters across multiple agents, and consequently, the parameters connected to a neuron are trained based on the gradients of all the agents who share that neuron if the neuron is shared. This enhances sample efficiency.  On the other hand, the parameters connected to a neuron are trained based only on  the gradient of one agent if only one agent owns the neuron. 
Thus, the obtained subnetworks as such allow multiple agents to learn common, locally common, and individual features when they are assigned to different agents. 
Such learning  can improve overall representational capacity by reinforcing agent indication compared with simple parameter sharing or parameter sharing with one-hot vector and also maintains sample efficiency through partial parameter sharing.   In addition, we can handle the amount of parameter-sharing across agents by adjusting the pruning ratio of each layer without introducing additional parameters. By controlling the pruning ratio, we can find  a soft spot achieving both high 
overall representational capacity and sample efficiency for high performance.
Note that such joint control of overall representational capacity and sample efficiency for MARL has not been achieved by the previous MARL parameter-sharing methods.
We summarized the procedure for generating winning group tickets based on the proposed structured network pruning method in Algorithm \ref{alg:snpps}, and Fig. \ref{fig:procedure} illustrates the overall procedure.

After obtaining the winning group tickets, we start training based on a MARL algorithm. Note that we need only one dense network to represent all $N$ subnetworks. With the assumption of centralized training with decentralized execution (CTDE), the gradients of the parameters are averaged over multiple agents. As aforementioned, this allows $N$-subnetworks to learn both common and individual features.

\begin{figure*}[t]
\begin{center}
\begin{tabular}{ccccc}
     % uncomment the next lines, and give the right ps files
     %\includegraphics[width=0.23\textwidth]{figures/MW_N2.png} &
     \includegraphics[width=0.2\textwidth]{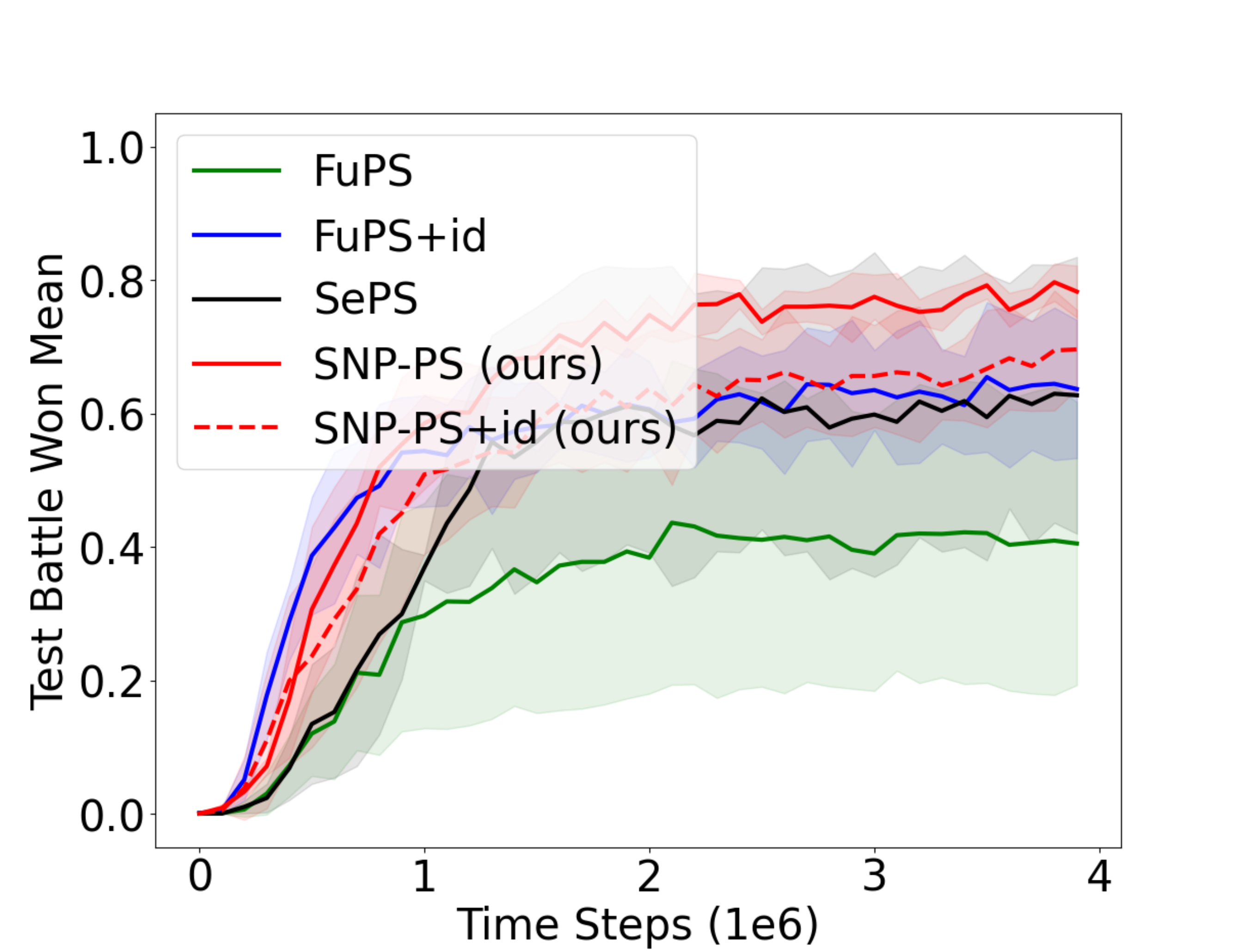} & \hspace{-3ex}
     \includegraphics[width=0.21\textwidth]{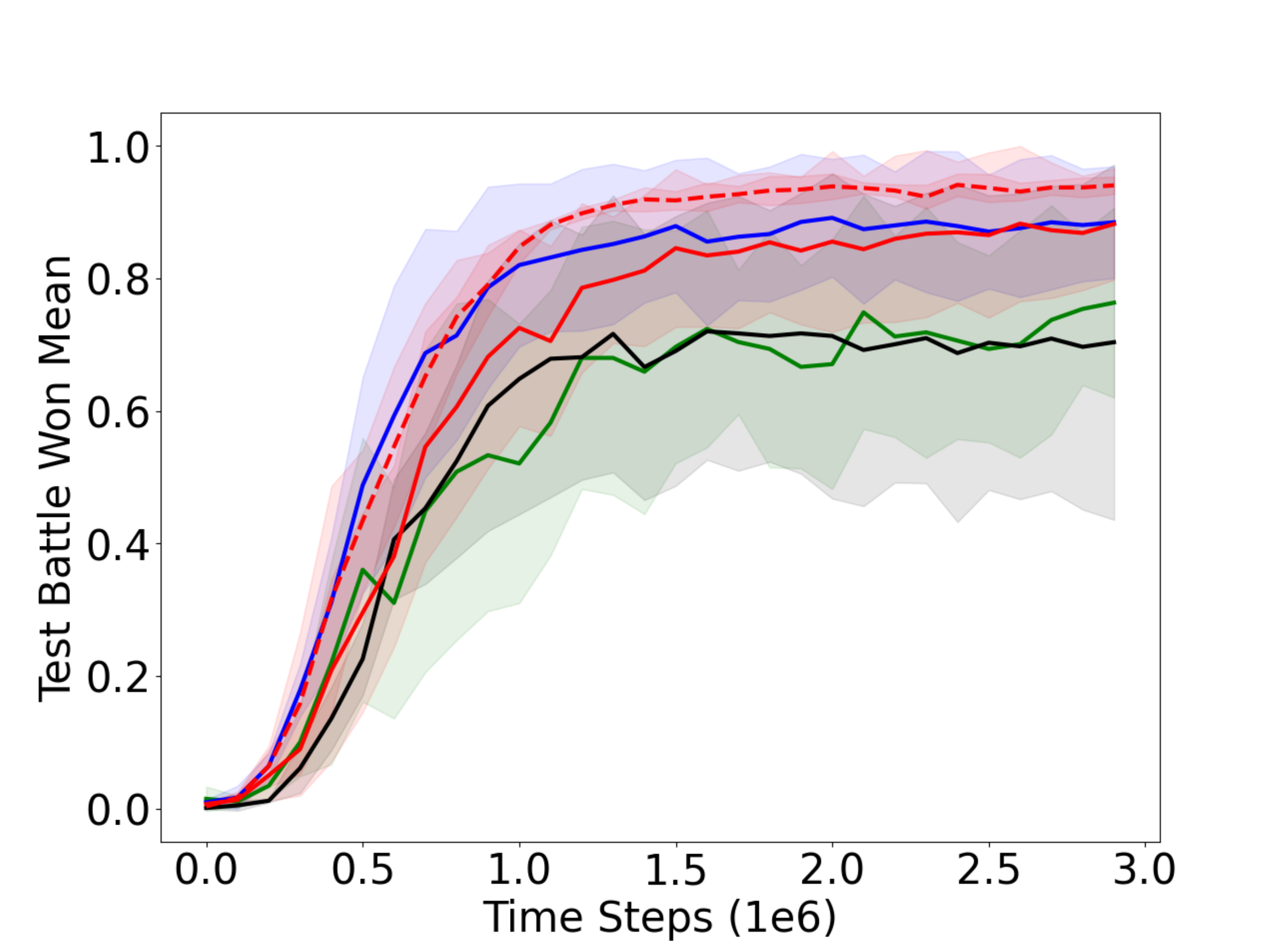} &
     \hspace{-3ex}
     \includegraphics[width=0.19\textwidth]{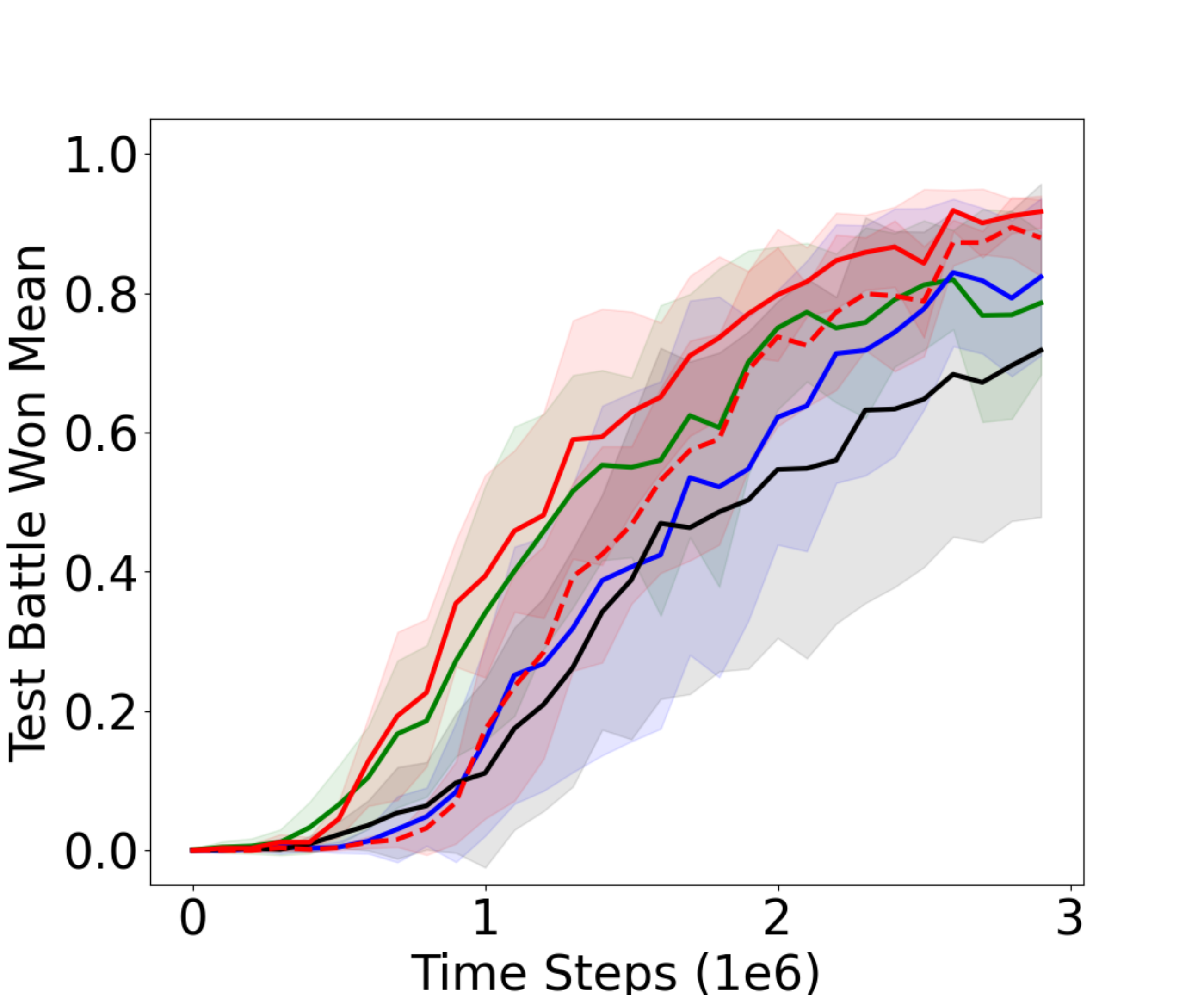} &
     \hspace{-3ex}
    \includegraphics[width=0.195\textwidth]{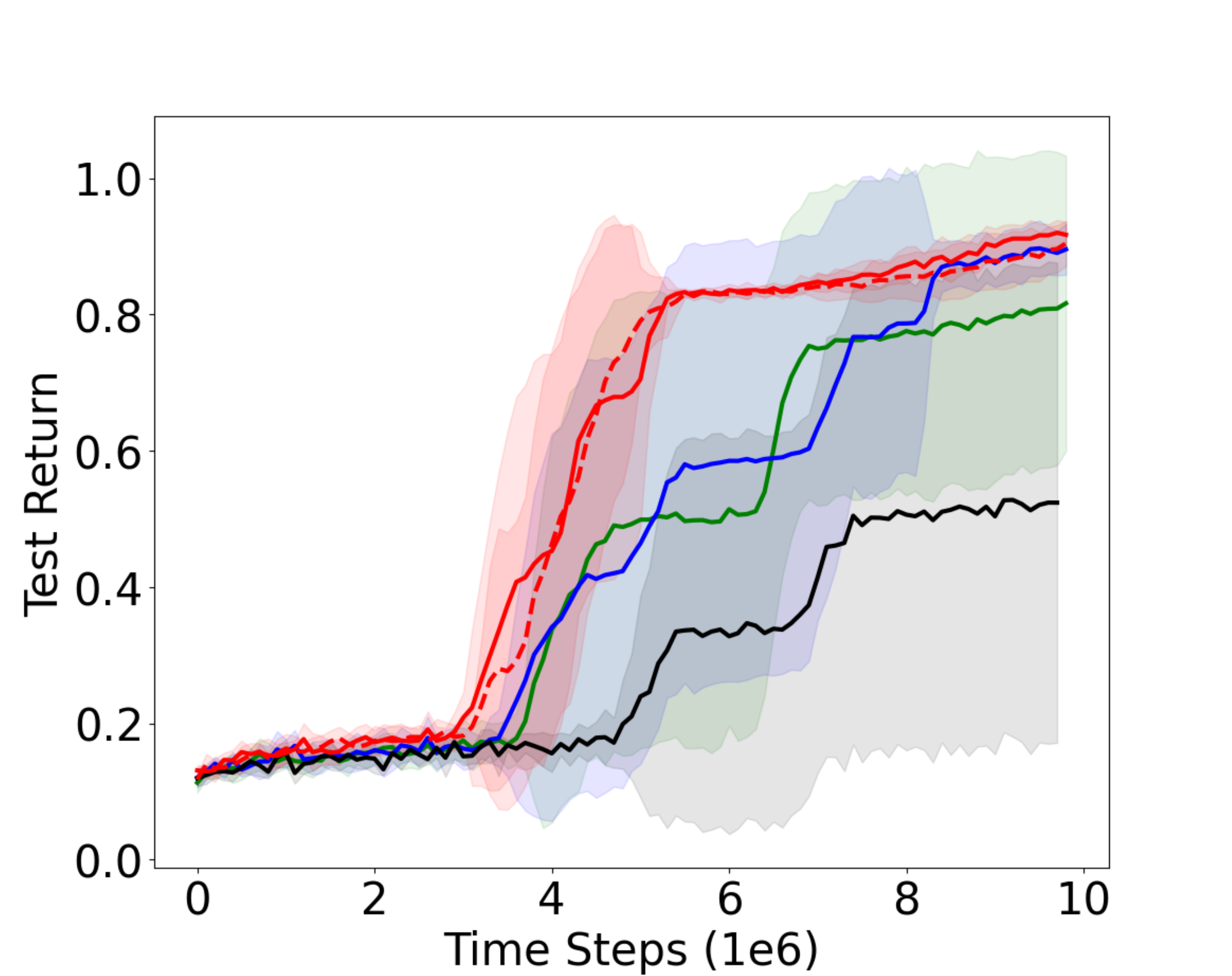} &
     \hspace{-3ex}
    \includegraphics[width=0.2\textwidth]{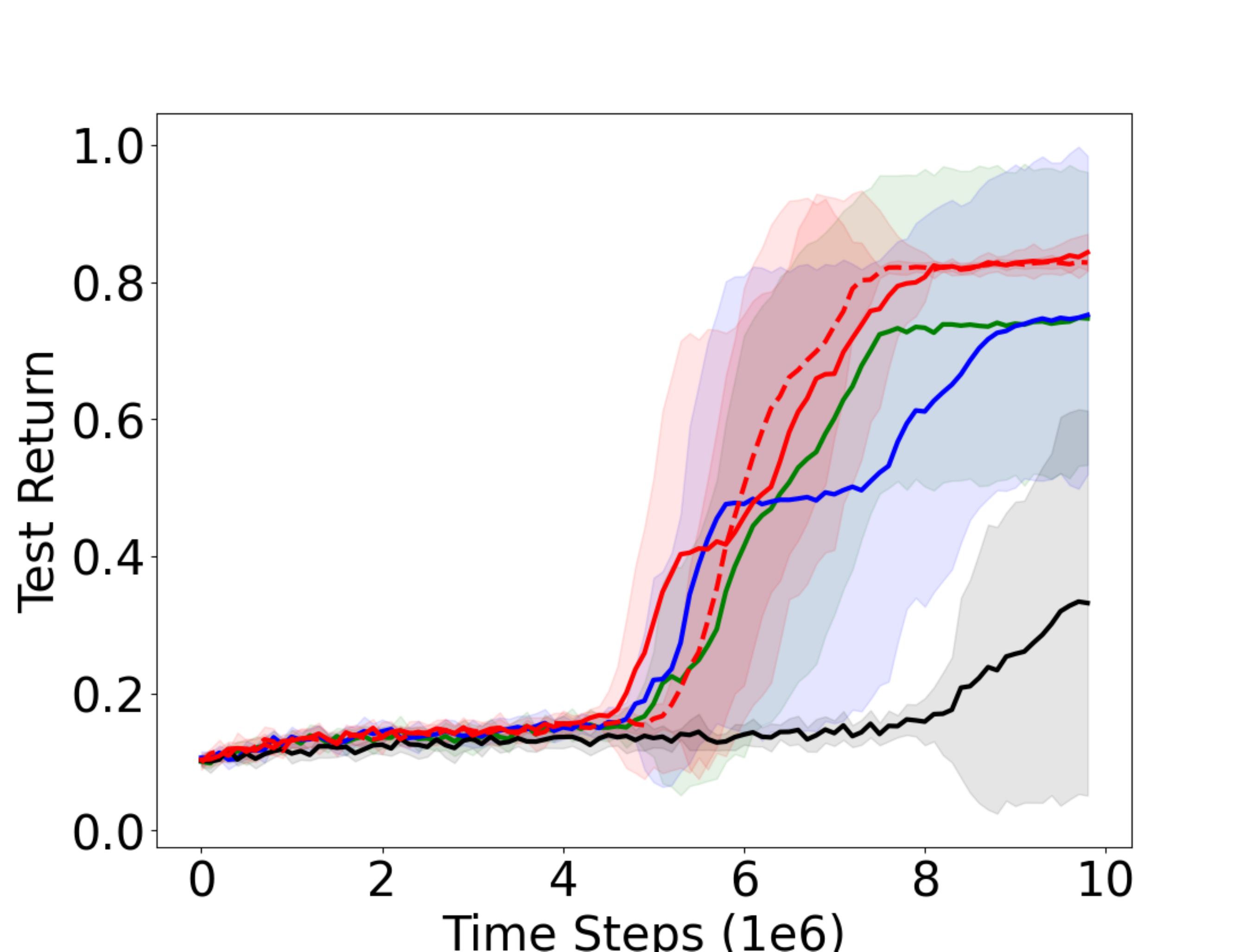}\\
     (a) 5m vs 6m (SMAC) &\hspace{-3ex} (b) 8m vs 9m (SMAC) &\hspace{-3ex} (c) MMM2 (SMAC) &\hspace{-3ex} (d) LBF1 &\hspace{-3ex} (e) LBF2  \\
\end{tabular}
\caption{Performance of FuPS (green), FuPS+id (blue), SePS (black), SNP-PS (red, ours), and SNP-PS+id (red-dotted, ours) on SMAC environments and LBF environments. More results are provided in Appendix.}
\vspace{-1em}
\label{fig:results}
\end{center}
\end{figure*}

\begin{table}[b]
  \caption{Environment information and pruning ratio of the proposed method in each environment. The pruning ratio '0-0.2-0.6' means that  the ratios of pruned neurons in the first, second, and third layers are $0\%$, $20\%$, and $60\%$, respectively. In the LBF environments, we used the same pruning ratio for actors and critics. A and C denote actors and critics, respectively}
  \label{table:env}
  \begin{tabular}{rccc}\toprule
    \textit{Environment} & \textit{$\#$ Agents} & \textit{$\#$ Type}  & \textit{Pruning ratio} \\ \midrule
    \textit{5m vs 6m} & 5 & 1 & 0.1-0.1 \\
    \textit{8m vs 9m} & 8 & 1 & 0-0.1 \\
    \textit{27m vs 30m} & 27 & 1 & 0-0.1 \\
    \textit{MMM2} & 10 & 3 & 0-0.1 \\
    \textit{LBF1} & 6 & 2 & 0-0.1-0.9 (C) 0-0.1-0.1 (A)\\
    \textit{LBF2} & 6 & 3 & 0-0.1-0.9 (C) 0-0.1-0.1 (A)\\ \bottomrule
  \end{tabular}
\end{table}

\section{Experiments}

In order to compare the proposed parameter-sharing  method with other parameter-sharing methods, we implemented the proposed method and other  baselines on top of QMIX and multi-agent A2C \cite{rashid2018qmix, christianos2021scaling}. We considered three parameter-sharing baselines: 1) FuPS - simple full parameter sharing in which all agents share the same parameters without agent indication. 2) FuPS+id - full parameter sharing with one-hot encoding agent indication. 3) SePS \cite{christianos2021scaling} - parameter sharing within each  subgroup and the subgroups  are constructed based on a clustering algorithm. We also considered the combination of our SNP-PS and one-hot encoding agent indication, which is denoted as SNP-PS+id.

\subsection{Environments}

\textbf{Starcraft II} ~~~~ We evaluated the proposed method on the StarcraftII micromanagement (SMAC) environment, which is commonly used as a benchmark of Dec-POMDP \cite{samvelyan2019starcraft}. We consider four tasks in SMAC: \textit{5m vs 6m}, \textit{8m vs 9m}, \textit{27m vs 30m} and \textit{MMM2}. \textit{MMM2} is a heterogeneous task consisting of three different types of agents, whereas the others are homogeneous tasks. For these tasks, we evaluate all parameter-sharing algorithms on top of QMIX. QMIX originally uses parameter sharing with one-hot encoding for individual action-value functions. For a fair comparison, we replace the parameter-sharing method for the individual action-value functions with the considered methods. The individual action-value functions have two 64-dimensional hidden vectors with three layers consisting of a GRU layer and two fully-connected layers before and after the GRU layer. \\
\textbf{Level-based Foraging} ~~~~ We also evaluated the proposed parameter-sharing method on the level-based forging (LBF) environment \cite{albrecht2015game} in which  multiple agents forage  randomly generated foods on a two-dimensional grid. The agents and foods have their own level and the agents can forage a food if the sum of levels of the agents who are adjacent to the food is equal to or larger than the level of the food. The agents receive a reward when they forage a food. We consider two LBF tasks that have different levels of agent and food. LBF1 consists of six agents whose levels are $(1,1,1,2,2,2)$ and six foods whose level is $3$. LBF2 consists of six agents whose levels are $(1,1,2,2,3,3)$ and  six foods whose level is $4$.
The agent and food levels here were slightly modified from those in 
\cite{christianos2021scaling} to make a more difficult environment.  In this environment, we implemented the proposed method on top of the multi-agent A2C algorithm considered in \cite{christianos2021scaling}. We applied  the parameter-sharing methods including the proposed method to both actor and critic. The actor and the critic consist of three 128-dimensional hidden vectors with four fully connected layers.

Table \ref{table:env} provides the details regarding the environments and the pruning ratio of the proposed method which is a hyperparameter.

\begin{figure*}[t]
\begin{center}
\begin{tabular}{ccc}
     % uncomment the next lines, and give the right ps files
     %\includegraphics[width=0.23\textwidth]{figures/MW_N2.png} &
     \includegraphics[width=0.3\textwidth]{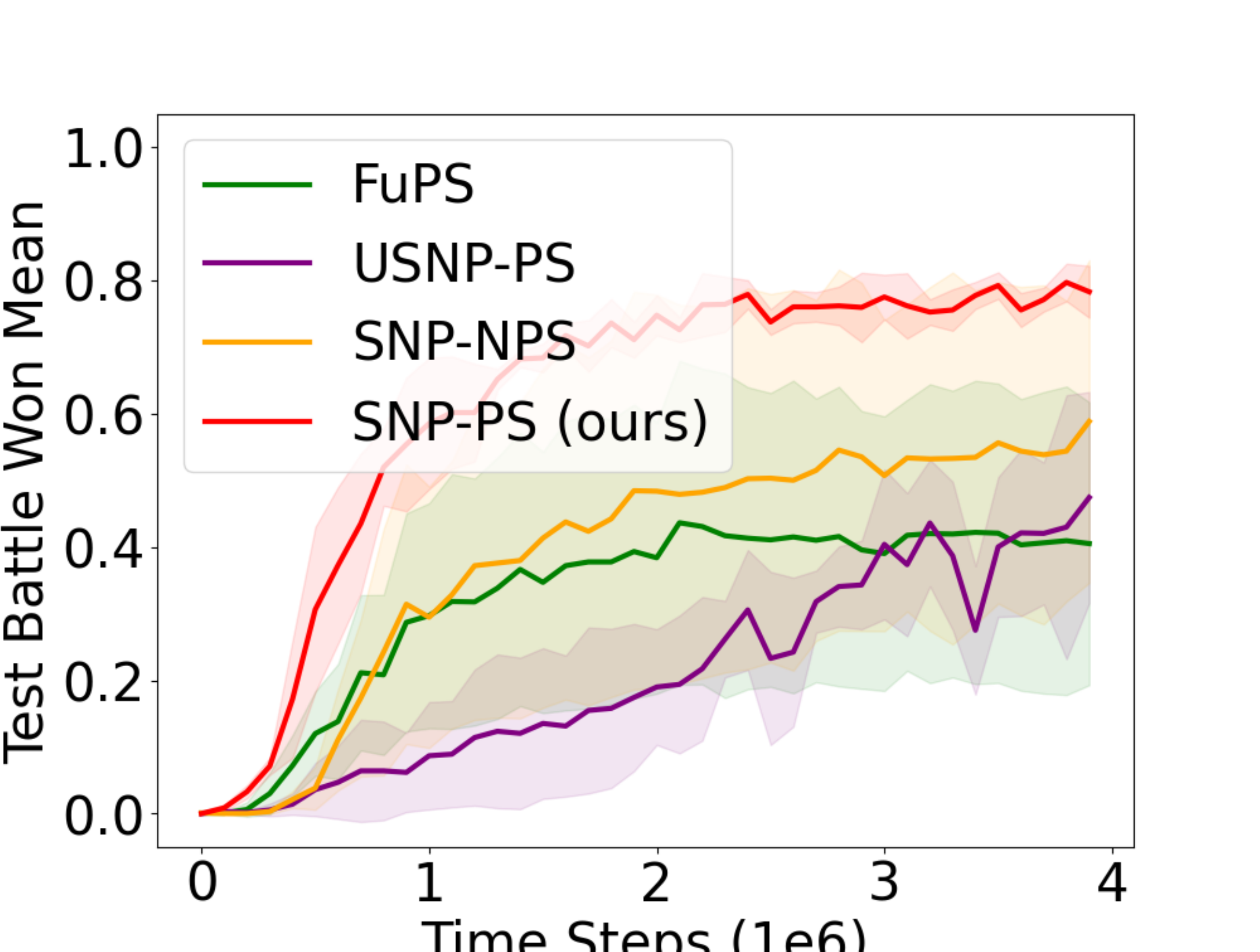} &
     \includegraphics[width=0.3\textwidth]{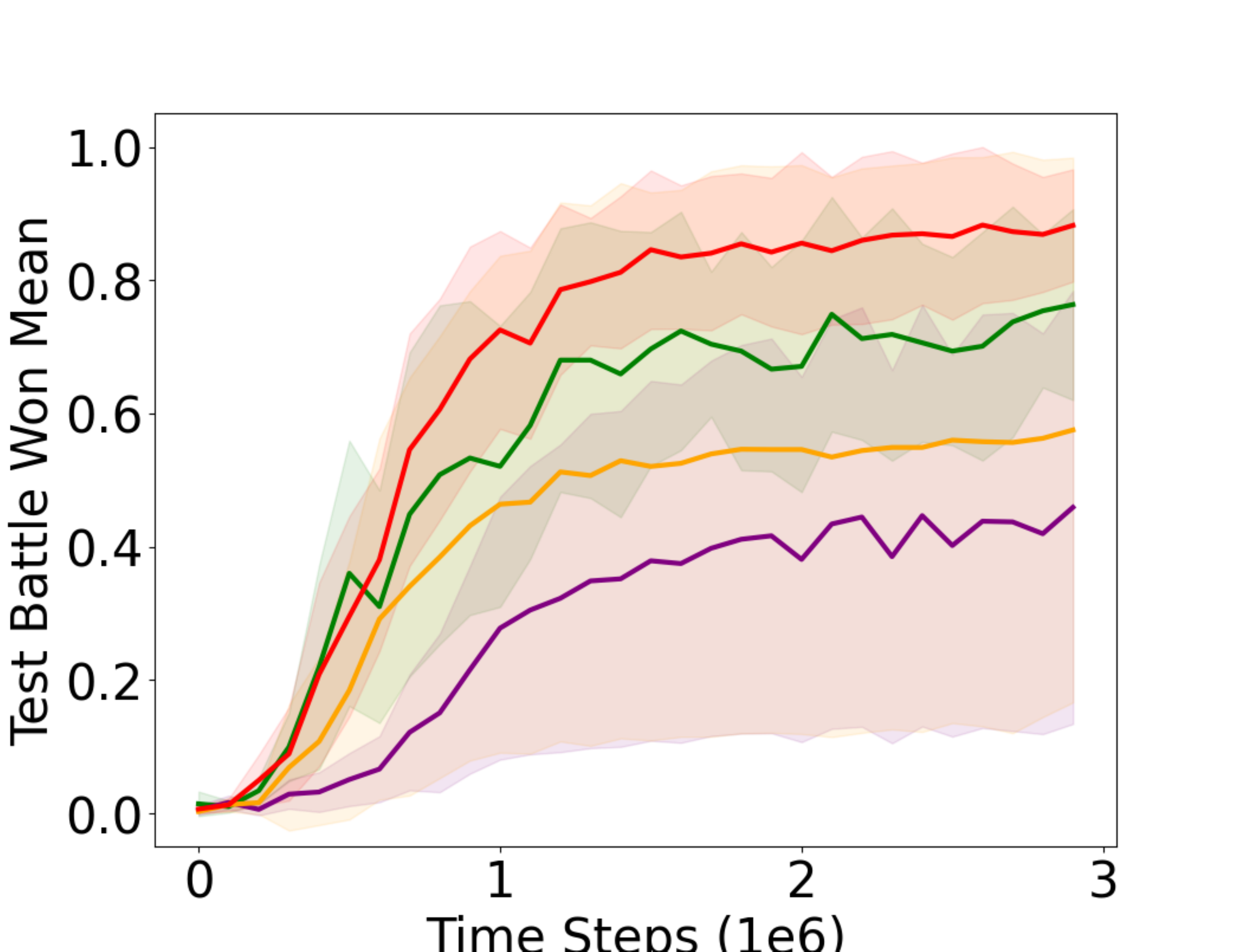} &
     \includegraphics[width=0.28\textwidth]{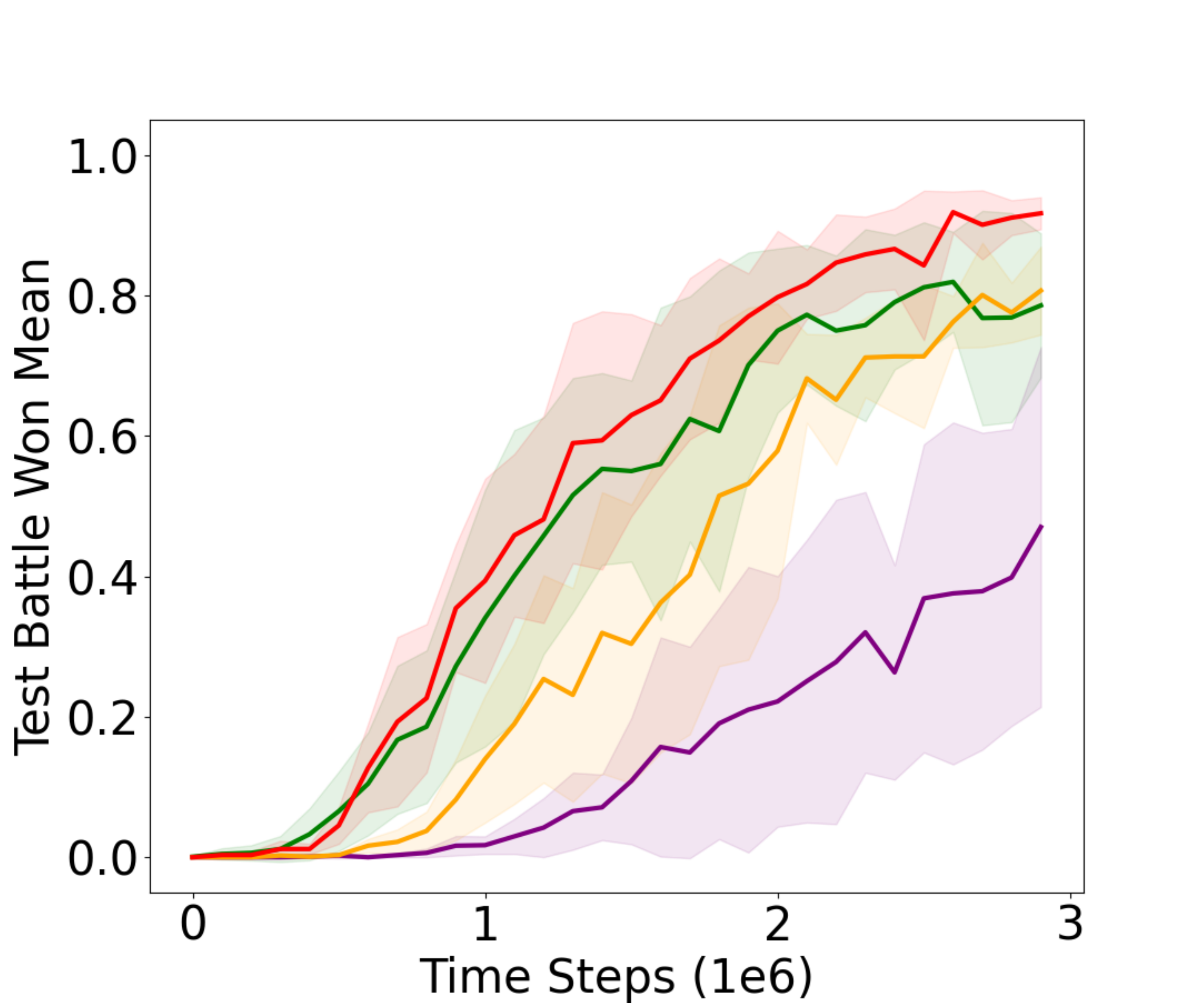}\\
     (a) 5m vs 6m & (b) 8m vs 9m & (c) MMM2 \\
\end{tabular}
 \caption{Comparison with other pruning methods: unstructured network pruning with parameter sharing (USNP-PS, purple) and structured network pruning with native parameter sharing (SNP-NPS, orange). USNP-PS removes the weights randomly while keeping the pruning ratio per layer. SNP-NPS generates one subnetwork based on the structured pruning method and then all agents share the subnetwork.}
\label{fig:othertickets}
\end{center}
\end{figure*}

\subsection{Results}\label{sec:results}

\textbf{Starcraft II} ~~~~ Figs. \ref{fig:results} (a)-(d) show the performance of the proposed method and the considered baselines on the SMAC environments. Parameter sharing with one-hot encoding agent indication (denoted as FuPS+id) performs better than simple  parameter sharing (denoted as FuPS) in \textit{5m vs 6m} and \textit{8m vs 9m} tasks, which consist of a relatively small number of agents. However, agent indication based on one-hot encoding does not improve performance in the \textit{27m vs 30m}  task consisting of many agents and the heterogeneous task (MMM2), and even degrades the training speed. SePS improves performance in \textit{5m vs 6m} compared with FuPS, but performs equally or worse in the other tasks. It seems that the sample inefficiency of SePS degrades the performance. The proposed method outperforms all the considered baselines. Especially in \textit{5m vs 6m}, which requires high-quality coordination, the performance improvement of our SNP-PS is significant due to the high representational capacity of the proposed method, resulting in diverse behaviors. \\
\textbf{Level-based Foraging} ~~~~ Figs.  \ref{fig:results} (e)-(f) show the performance of the proposed method and the considered baselines on the LBF environments. In terms of the final performance and training speed, the proposed method outperforms the considered baselines.

\subsection{Analysis: Construction of the Winning Group Tickets}

We proposed using randomly neuron-pruned subnetworks as winning group tickets for the lottery group ticket hypothesis. As seen in Fig. \ref{fig:results}, the proposed method outperforms the simple parameter sharing, which implies it succeeds in finding  winning group tickets. To see the effectiveness of structured pruning, we examine whether other pruning methods such as unstructured network pruning can obtain  winning group tickets or not.

\textbf{Unstructured pruning.} ~~~~~ We compared our structured network pruning method with the unstructured pruning method (denoted as USNP-PS in Fig. \ref{fig:othertickets}, which removes the weights randomly with the same pruning ratio. As seen in Fig. \ref{fig:othertickets}, the unstructured pruning method performs worse than the simple parameter sharing, which implies it fails to find winning group tickets.

\textbf{Parameter sharing with one pruned subnetwork.} ~~~~~ The proposed pruning method generates $N$ different subnetworks that come from a root network. It enables agents to share parts of the parameters. In order to see if performance improvement comes from the use of the pruned subnetwork, we compared the proposed method with the simple parameter sharing based on one selected pruned subnetwork, which is denoted as SNP-NPS in Fig. \ref{fig:othertickets}. That is, SNP-NPS generates one subnetwork based on the structured pruning method instead of generating $N$ subnetworks independently, and then all agents  use the same pruned network for naive parameter sharing. As seen in Fig. \ref{fig:othertickets}, SNP-NPS is not a winning group ticket.
It is seen that  the two considered methods are not an appropriate way to find winning group tickets, but the proposed method is an adequate way to construct winning group tickets  outperforming the naive parameter-sharing method in the considered environments.

\subsection{Analysis: Pruning Ratio}

We provided the pruning ratio we used in Table \ref{table:env}.  In the SMAC environment, we found that pruning the first hidden neurons performs better in hard tasks such as \textit{5m vs 6m}, but keeping all neurons in the first hidden neurons performs better in the other relatively easy environments. In the LBF environment, we conducted experiments to show the effect of the pruning ratio on the actor and critic networks, respectively.

\begin{figure}[t]
\begin{center}
\begin{tabular}{cc}
     % uncomment the next lines, and give the right ps files
     %\includegraphics[width=0.23\textwidth]{figures/MW_N2.png} &
     \includegraphics[width=0.25\textwidth]{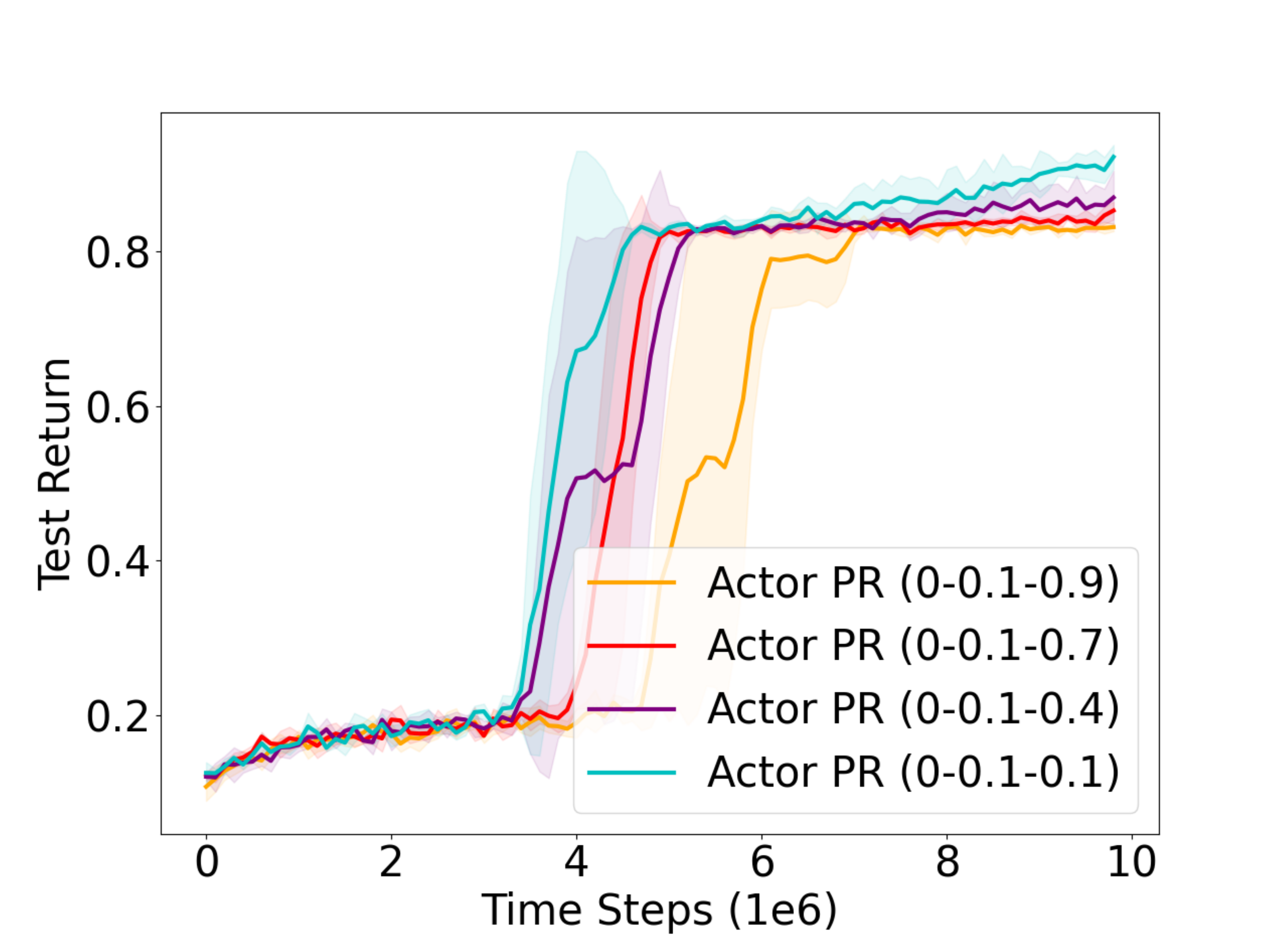} \hspace{-4ex} &
     \includegraphics[width=0.24\textwidth]{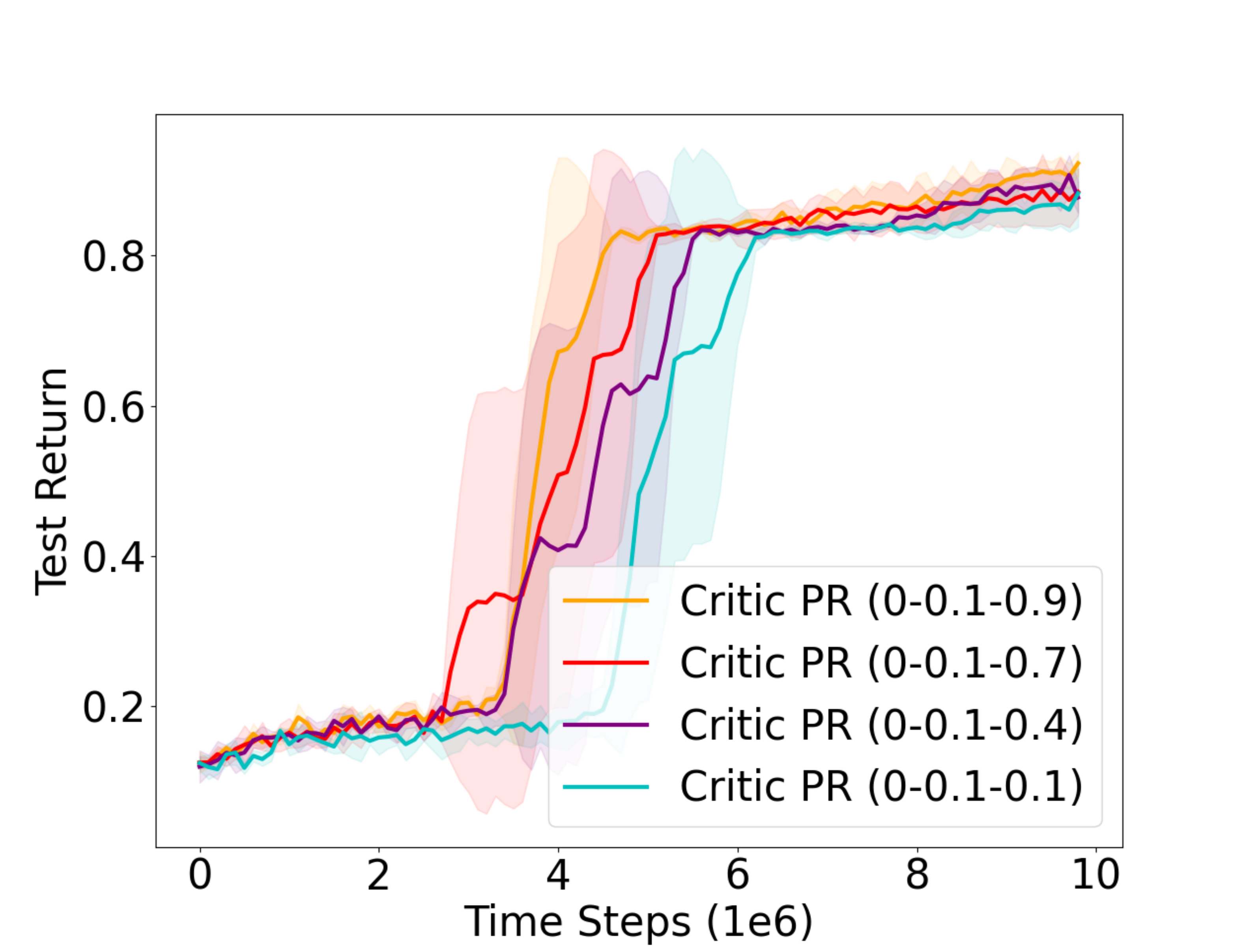}\\
     (a)  \hspace{-4ex} & (b) \\
\end{tabular}
 \caption{Performances with respect to the pruning ratio (PR) of actor networks and critics networks. (a) the PR of critic networks is fixed as '0-0.1-0.9' and that of the last hidden vector in actor networks varies from $0.1$ to $0.9$. (b) the PR of actor networks is fixed as '0-0.1-0.1' and that of the last hidden vector in critic networks varies from $0.1$ to $0.9$.}
\label{fig:pruningratio}
\end{center}
\end{figure}

\textbf{The effect of pruning on actor network.}  ~~~~~ We conducted experiments by fixing the pruning ratio of the critic neural network and changing that of the actor neural network. The result is shown in Fig. \ref{fig:pruningratio} (a). It is seen that the performance is the highest when the pruning ratio is the smallest, but the steady-state performance is not sensitive to the pruning ratio of the actor neural network.

%It is seen that the training speed is the lowest when the pruning ratio is the smallest, but the performance is not sensitive to the pruning ratio of the actor neural network.

\textbf{The effect of pruning on critic network.}  ~~~~~ We conducted experiments by fixing the pruning ratio of the actor neural network and changing that of the critic neural network. The result is shown in Fig.  \ref{fig:pruningratio} (b). It is seen that  the training speed and the final performance increase as the pruning ratio of the critic network increases. Since agents are heterogeneous and receive individual rewards in the LBF environment, each agent should estimate its own return and thus the critic should be more identifiable. Highly identifiable critics can be obtained by increasing the pruning ratio of the critic network.

Note that the best pruning ratio can vary across environments and thus we leave how to choose the proper pruning ratio as future work.

\begin{figure*}[t]
\begin{center}
\begin{tabular}{cccc}
     % uncomment the next lines, and give the right ps files
     %\includegraphics[width=0.23\textwidth]{figures/MW_N2.png} &
     \includegraphics[width=0.22\textwidth]{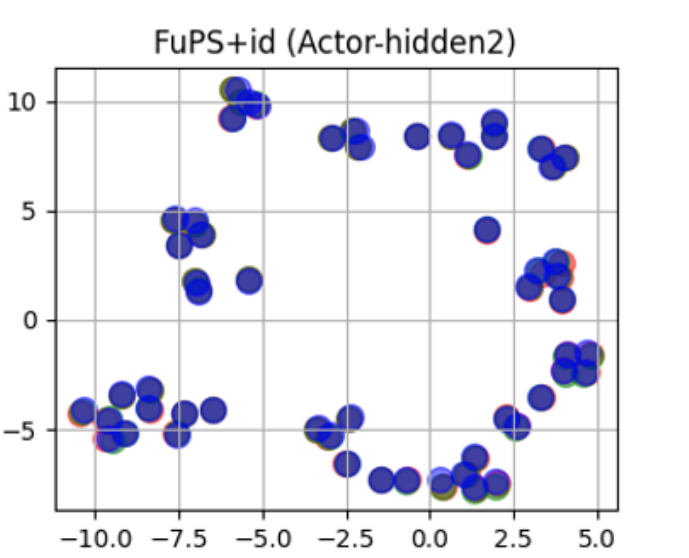} &
     \includegraphics[width=0.22\textwidth]{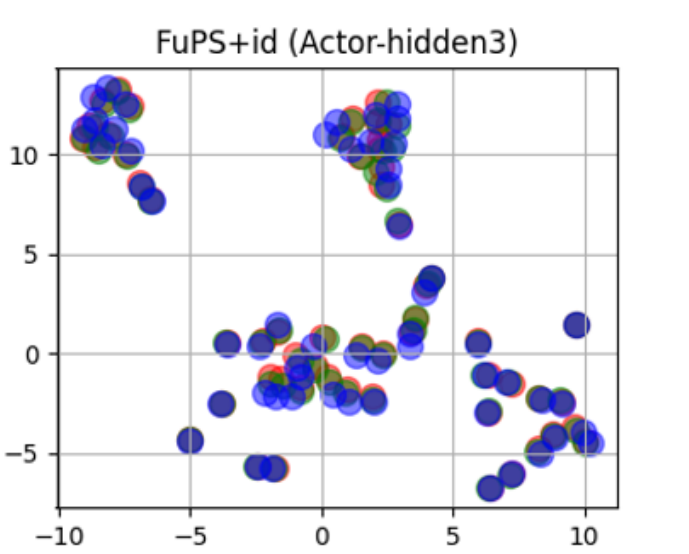} &
     \includegraphics[width=0.22\textwidth]{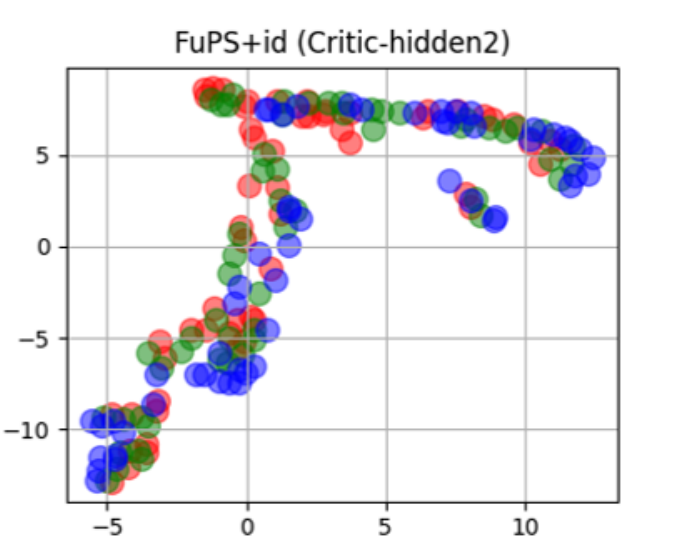} &
     \includegraphics[width=0.22\textwidth]{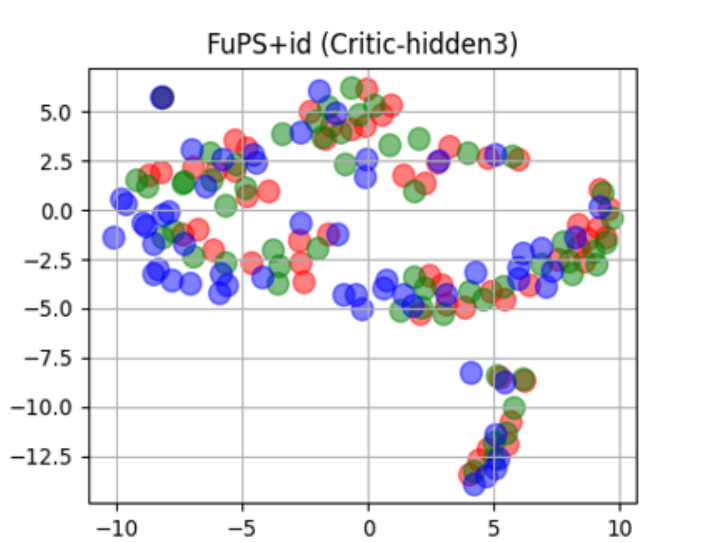}\\
     \includegraphics[width=0.22\textwidth]{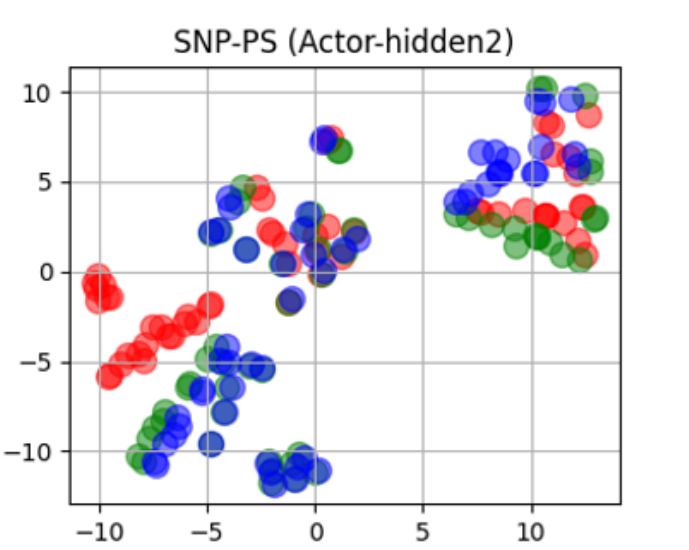} &
     \includegraphics[width=0.22\textwidth]{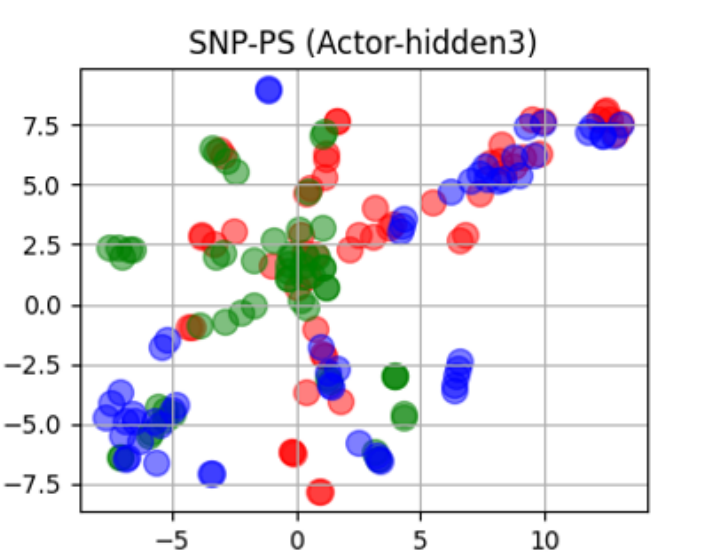} &
     \includegraphics[width=0.22\textwidth]{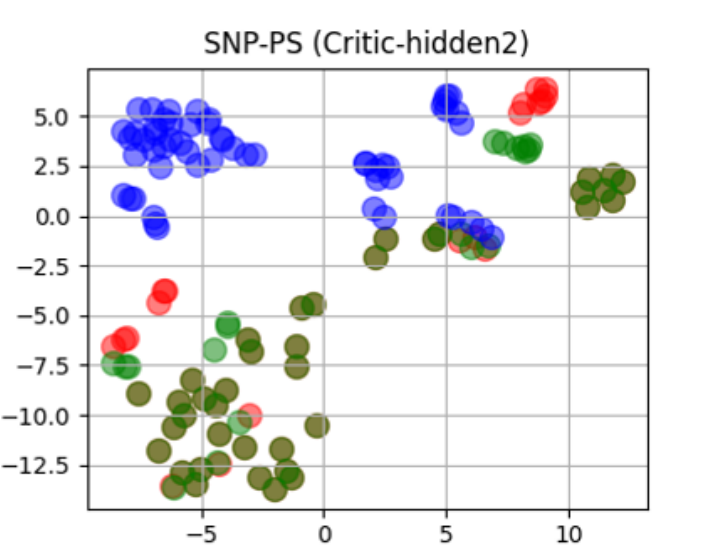} &
     \includegraphics[width=0.22\textwidth]{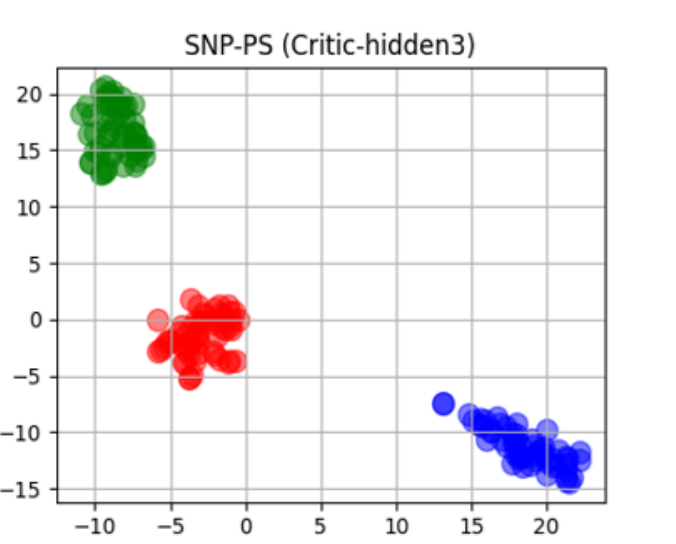}\\
\end{tabular}
\caption{T-SNE plot of hidden features: 1st row - FuPS+id and 2nd row - SNP-PS. Each column corresponds to the result of 2nd hidden feature of actor, 3rd hidden feature of actor, 2nd hidden feature of critic, and 3rd hidden feature of critic, respectively. We consider  three agents in the LBF1 environment, with the blue, red, and green circles denoting each respective agent.}
\label{fig:results_t_sne}
\end{center}
\end{figure*}

\subsection{Analysis: Hidden Representations}

The proposed method allows agents to perform different actions even with the same input by increasing the representational capacity of naive parameter sharing. To analyze this, we visualize the hidden features of three agents' networks with the same observation. 
Note that native parameter sharing yields identical hidden features for all agents given the same observation. We here investigate the hidden features generated by the proposed method and the one-hot encoding agent indication, as shown in Fig. \ref{fig:results_t_sne}. The hidden features of multiple agents generated by the one-hot encoding agent indication display a significant degree of overlap or minimal variation, whereas those generated by the proposed method comprise both partially overlapped features and non-overlapping features, which shows the agents with the same observation have either distinct or similar hidden representation. Thus, in Fig. \ref{fig:results_t_sne}, it is shown that the proposed method has a high degree of representational capacity compared to the one-hot encoding agent indication.

\subsection{Analysis: Resource}

In  real-world decision-making problems, environmental resources such as samples, memory size, and computational cost are constrained. For the applicability of MARL algorithms to practical problems, MARL algorithms should be able to perform in resource-constrained environments. We investigate the considered algorithms from the perspective of resource constraints.

\textbf{Sample efficiency. }  ~~~~~ Sample inefficiency is one notorious limitation of RL, and the problem aggravates in MARL due to increasing variance for gradient estimation. Since the proposed method learns the common and individual features effectively, it is seen that the training speed of the proposed method is the highest among the considered parameter-sharing algorithms. Thus, the proposed method performs better than other baselines in terms of sample efficiency.

\textbf{Memory size.} ~~~~~ Without a parameter-sharing method, the required memory size increases linearly as the number of agents increases. Parameter-sharing methods address this problem, but one-hot encoding agent indication and SePS still need more parameters than simple parameter sharing. However, the proposed method requires the same number of parameters as the simple parameter sharing and learns the common and individual features based on a structured pruning method. Table \ref{table:resource} shows the required number of parameters for the considered algorithms in the LBF environment.

\begin{figure}[h]
\begin{center}
\begin{tabular}{c}
     % uncomment the next lines, and give the right ps files
     %\includegraphics[width=0.23\textwidth]{figures/MW_N2.png} &
     \includegraphics[width=0.25\textwidth]{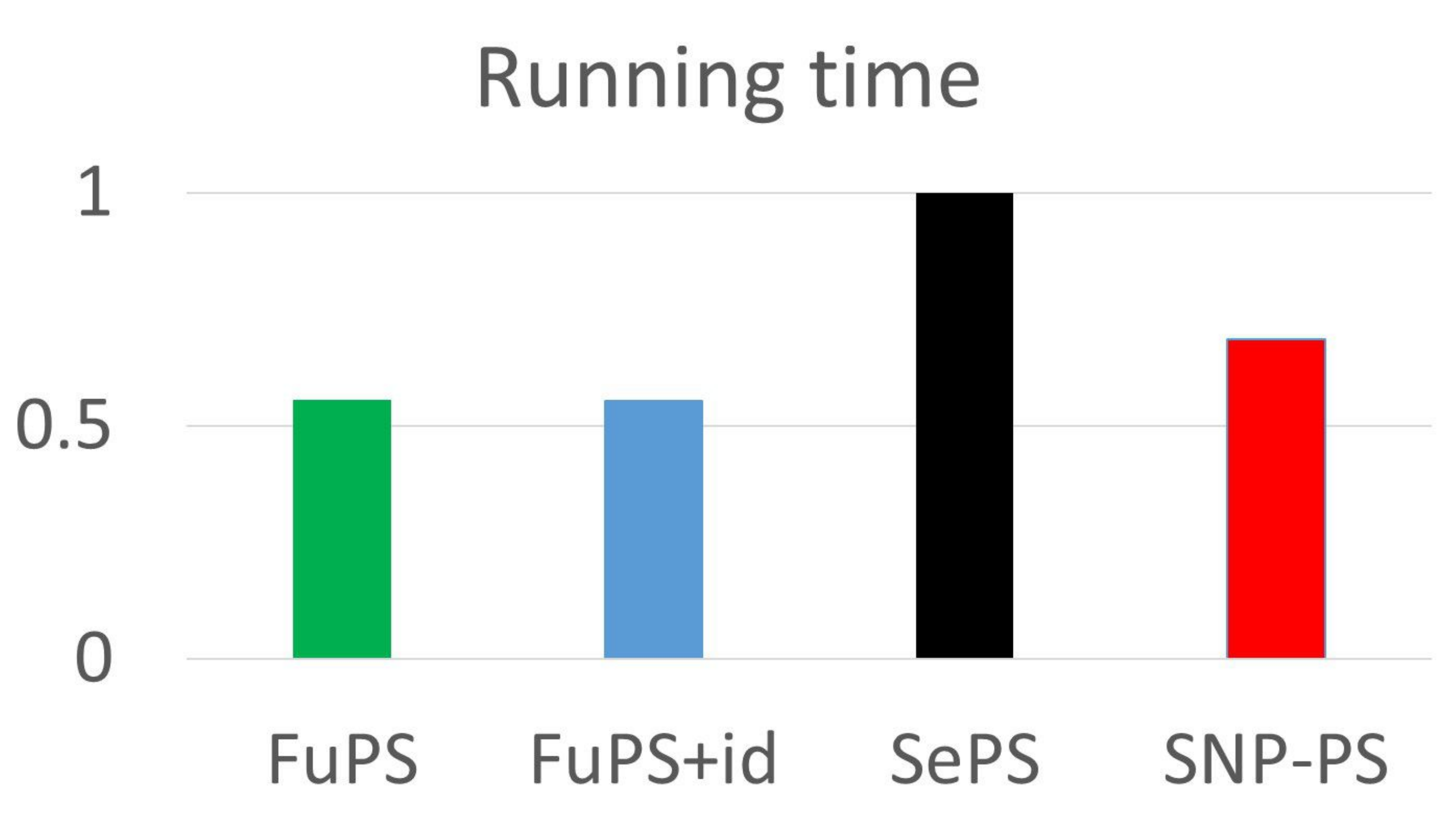} \\
     %(a) \vspace{2ex} \\ 
     %\includegraphics[width=0.14\textwidth]{figures/onehot1.pdf} \hspace{1ex}
     %\includegraphics[width=0.14\textwidth]{figures/onehot2.pdf} \hspace{1ex}
     %\includegraphics[width=0.14\textwidth]{figures/onehot3.pdf}\\
     %\hspace{-4ex} (b) \hspace{17ex} (c) \hspace{17ex} (d) 
\end{tabular}
 \caption{Normalized running time during training.}
\label{fig:runningtime}
\end{center}
\end{figure}

\textbf{Computational cost.} ~~~~~ We provided the running time during training normalized by the maximum value in Fig. \ref{fig:runningtime}.  Since the proposed method computes the multiplication of the hidden vector and the generated binary mask, it requires  slightly more computation compared with the naive parameter sharing as seen in Fig. \ref{fig:runningtime}. However, multiplication with binary mask is trivial. The proposed method has a lower computational cost than SePS, which uses a clustering algorithm and several neural networks for parameterizing the actor and critic.

\begin{table}[t]
  \caption{The required number of parameters in the LBF environment. We consider $K=3$ clusters for SePS.}
  \label{table:resource}
  \begin{tabular}{rcccc}\toprule
    \textit{Algorithm} & \textit{FuPS} & \textit{FuPS+id} & \textit{SePS} & \textit{SNP-PS} \\ \midrule
    \textit{$\#$ parameters} & \textbf{76k} & 78k & 229k & \textbf{76k}\\ \bottomrule
  \end{tabular}
\end{table}

\begin{comment}
\begin{table}[h]
  \caption{The required number of parameters in the proposed method and the considered baselines}
  \label{table:resource}
  \begin{tabular}{rccc}\toprule
    \textit{$\#$ parameters} & \textit{SNP-PS} \\ \midrule
    \textit{5m vs 6m} & 5 \\
    \textit{8m vs 9m} & 8 \\
    \textit{27m vs 30m} & \\
    \textit{MMM2} & 10  \\
    \textit{LBF1} & 6 \\
    \textit{LBF2} & 6\\ \bottomrule
  \end{tabular}
\end{table}
\end{comment}

\section{Conclusion}

We have proposed a simple yet effective technique for parameter sharing for multi-agent deep reinforcement learning to improve the representational capacity of the joint policy and/or critic while keeping high sample efficiency. The proposed method generates multiple subnetworks  for multi-agent policies and/or critics by randomly and structurally pruning a common dense network 
without introducing additional parameters.  Subnetworks possess overlapping shared parameters and individual parameters. Due to this mixture property, the proposed structured pruning-based parameter-sharing method achieves both high joint representational capacity and sample efficiency with a proper pruning ratio.  We have evaluated the proposed method on several benchmark tasks, and numerical results show that the proposed method noticeably outperforms other parameter-sharing methods.  The practical effectiveness of the proposed method is verified by experiments but theoretical proof of the lottery ticket hypothesis and lottery group ticket hypothesis remains, which is a meaningful work in general deep learning theory.

\section{Acknowledgements}
This work was partly supported by Institute of Information $\&$ communications Technology Planing $\&$ Evaluation (IITP) grant funded by the Korea government(MSIT) (No.2022-0-00124), Development of Artificial Intelligence Technology for Self-Improving Competency-Aware Learning Capabilities) and the National Research Foundation of Korea(NRF) grant funded by the Korea government(MSIT). (NRF-2021R1A2C2009143)

%%%%%%%%%%%%%%%%%%%%%%%%%%%%%%%%%%%%%%%%%%%%%%%%%%%%%%%%%%%%%%%%%%%%%%%%

%%% The next two lines define, first, the bibliography style to be 
%%% applied, and, second, the bibliography file to be used.

\bibliographystyle{ACM-Reference-Format} 
\bibliography{main}


\begin{thebibliography}{23}

%%% the Web address.

\ifx \showCODEN    \undefined \def \showCODEN     #1{\unskip}     \fi
\ifx \showDOI      \undefined \def \showDOI       #1{#1}\fi
\ifx \showISBNx    \undefined \def \showISBNx     #1{\unskip}     \fi
\ifx \showISBNxiii \undefined \def \showISBNxiii  #1{\unskip}     \fi
\ifx \showISSN     \undefined \def \showISSN      #1{\unskip}     \fi
\ifx \showLCCN     \undefined \def \showLCCN      #1{\unskip}     \fi
\ifx \shownote     \undefined \def \shownote      #1{#1}          \fi
\ifx \showarticletitle \undefined \def \showarticletitle #1{#1}   \fi
\ifx \showURL      \undefined \def \showURL       {\relax}        \fi
% invisible to TeX
\providecommand\bibfield[2]{#2}
\providecommand\bibinfo[2]{#2}
\providecommand\natexlab[1]{#1}
\providecommand\showeprint[2][]{arXiv:#2}

\bibitem[\protect\citeauthoryear{Albrecht and Ramamoorthy}{Albrecht and
  Ramamoorthy}{2015}]%
        {albrecht2015game}
\bibfield{author}{\bibinfo{person}{Stefano~V Albrecht} {and}
  \bibinfo{person}{Subramanian Ramamoorthy}.} \bibinfo{year}{2015}\natexlab{}.
\newblock \showarticletitle{A game-theoretic model and best-response learning
  method for ad hoc coordination in multiagent systems}.
\newblock \bibinfo{journal}{\emph{arXiv preprint arXiv:1506.01170}}
  (\bibinfo{year}{2015}).
\newblock


\bibitem[\protect\citeauthoryear{Christianos, Papoudakis, Rahman, and
  Albrecht}{Christianos et~al\mbox{.}}{2021}]%
        {christianos2021scaling}
\bibfield{author}{\bibinfo{person}{Filippos Christianos},
  \bibinfo{person}{Georgios Papoudakis}, \bibinfo{person}{Muhammad~A Rahman},
  {and} \bibinfo{person}{Stefano~V Albrecht}.} \bibinfo{year}{2021}\natexlab{}.
\newblock \showarticletitle{Scaling multi-agent reinforcement learning with
  selective parameter sharing}. In \bibinfo{booktitle}{\emph{International
  Conference on Machine Learning}}. PMLR, \bibinfo{pages}{1989--1998}.
\newblock


\bibitem[\protect\citeauthoryear{Foerster, Assael, De~Freitas, and
  Whiteson}{Foerster et~al\mbox{.}}{2016}]%
        {foerster2016learning}
\bibfield{author}{\bibinfo{person}{Jakob Foerster},
  \bibinfo{person}{Ioannis~Alexandros Assael}, \bibinfo{person}{Nando
  De~Freitas}, {and} \bibinfo{person}{Shimon Whiteson}.}
  \bibinfo{year}{2016}\natexlab{}.
\newblock \showarticletitle{Learning to communicate with deep multi-agent
  reinforcement learning}.
\newblock \bibinfo{journal}{\emph{Advances in neural information processing
  systems}}  \bibinfo{volume}{29} (\bibinfo{year}{2016}).
\newblock


\bibitem[\protect\citeauthoryear{Foerster, Farquhar, Afouras, Nardelli, and
  Whiteson}{Foerster et~al\mbox{.}}{2018}]%
        {foerster2018counterfactual}
\bibfield{author}{\bibinfo{person}{Jakob Foerster}, \bibinfo{person}{Gregory
  Farquhar}, \bibinfo{person}{Triantafyllos Afouras}, \bibinfo{person}{Nantas
  Nardelli}, {and} \bibinfo{person}{Shimon Whiteson}.}
  \bibinfo{year}{2018}\natexlab{}.
\newblock \showarticletitle{Counterfactual multi-agent policy gradients}. In
  \bibinfo{booktitle}{\emph{Proceedings of the AAAI conference on artificial
  intelligence}}, Vol.~\bibinfo{volume}{32}.
\newblock


\bibitem[\protect\citeauthoryear{Frankle and Carbin}{Frankle and
  Carbin}{2018}]%
        {frankle2018lottery}
\bibfield{author}{\bibinfo{person}{Jonathan Frankle} {and}
  \bibinfo{person}{Michael Carbin}.} \bibinfo{year}{2018}\natexlab{}.
\newblock \showarticletitle{The Lottery Ticket Hypothesis: Finding Sparse,
  Trainable Neural Networks}. In \bibinfo{booktitle}{\emph{International
  Conference on Learning Representations}}.
\newblock


\bibitem[\protect\citeauthoryear{Graesser, Evci, Elsen, and Castro}{Graesser
  et~al\mbox{.}}{2022}]%
        {graesser2022state}
\bibfield{author}{\bibinfo{person}{Laura Graesser}, \bibinfo{person}{Utku
  Evci}, \bibinfo{person}{Erich Elsen}, {and} \bibinfo{person}{Pablo~Samuel
  Castro}.} \bibinfo{year}{2022}\natexlab{}.
\newblock \showarticletitle{The State of Sparse Training in Deep Reinforcement
  Learning}. In \bibinfo{booktitle}{\emph{International Conference on Machine
  Learning}}. PMLR, \bibinfo{pages}{7766--7792}.
\newblock


\bibitem[\protect\citeauthoryear{Iqbal and Sha}{Iqbal and Sha}{2019}]%
        {iqbal2019actor}
\bibfield{author}{\bibinfo{person}{Shariq Iqbal} {and} \bibinfo{person}{Fei
  Sha}.} \bibinfo{year}{2019}\natexlab{}.
\newblock \showarticletitle{Actor-attention-critic for multi-agent
  reinforcement learning}. In \bibinfo{booktitle}{\emph{International
  conference on machine learning}}. PMLR, \bibinfo{pages}{2961--2970}.
\newblock


\bibitem[\protect\citeauthoryear{Jeon, Kim, Jung, and Sung}{Jeon
  et~al\mbox{.}}{2022}]%
        {jeon2022maser}
\bibfield{author}{\bibinfo{person}{Jeewon Jeon}, \bibinfo{person}{Woojun Kim},
  \bibinfo{person}{Whiyoung Jung}, {and} \bibinfo{person}{Youngchul Sung}.}
  \bibinfo{year}{2022}\natexlab{}.
\newblock \showarticletitle{MASER: Multi-Agent Reinforcement Learning with
  Subgoals Generated from Experience Replay Buffer}. In
  \bibinfo{booktitle}{\emph{International Conference on Machine Learning}}.
  PMLR, \bibinfo{pages}{10041--10052}.
\newblock


\bibitem[\protect\citeauthoryear{Kim, Cho, and Sung}{Kim et~al\mbox{.}}{2019}]%
        {kim2019message}
\bibfield{author}{\bibinfo{person}{Woojun Kim}, \bibinfo{person}{Myungsik Cho},
  {and} \bibinfo{person}{Youngchul Sung}.} \bibinfo{year}{2019}\natexlab{}.
\newblock \showarticletitle{Message-dropout: An efficient training method for
  multi-agent deep reinforcement learning}. In
  \bibinfo{booktitle}{\emph{Proceedings of the AAAI conference on artificial
  intelligence}}, Vol.~\bibinfo{volume}{33}. \bibinfo{pages}{6079--6086}.
\newblock


\bibitem[\protect\citeauthoryear{Kim, Jung, Cho, and Sung}{Kim
  et~al\mbox{.}}{2023}]%
        {kim2023variational}
\bibfield{author}{\bibinfo{person}{Woojun Kim}, \bibinfo{person}{Whiyoung
  Jung}, \bibinfo{person}{Myungsik Cho}, {and} \bibinfo{person}{Youngchul
  Sung}.} \bibinfo{year}{2023}\natexlab{}.
\newblock \showarticletitle{A Variational Approach to Mutual Information-Based
  Coordination for Multi-Agent Reinforcement Learning}.
\newblock  (\bibinfo{year}{2023}).
\newblock
\showeprint[arxiv]{2303.00451}~[cs.MA]


\bibitem[\protect\citeauthoryear{Kim, Park, and Sung}{Kim
  et~al\mbox{.}}{2020}]%
        {kim2020communication}
\bibfield{author}{\bibinfo{person}{Woojun Kim}, \bibinfo{person}{Jongeui Park},
  {and} \bibinfo{person}{Youngchul Sung}.} \bibinfo{year}{2020}\natexlab{}.
\newblock \showarticletitle{Communication in multi-agent reinforcement
  learning: Intention sharing}. In \bibinfo{booktitle}{\emph{International
  Conference on Learning Representations}}.
\newblock


\bibitem[\protect\citeauthoryear{Livne and Cohen}{Livne and Cohen}{2020}]%
        {livne2020pops}
\bibfield{author}{\bibinfo{person}{Dor Livne} {and} \bibinfo{person}{Kobi
  Cohen}.} \bibinfo{year}{2020}\natexlab{}.
\newblock \showarticletitle{Pops: Policy pruning and shrinking for deep
  reinforcement learning}.
\newblock \bibinfo{journal}{\emph{IEEE Journal of Selected Topics in Signal
  Processing}} \bibinfo{volume}{14}, \bibinfo{number}{4}
  (\bibinfo{year}{2020}), \bibinfo{pages}{789--801}.
\newblock


\bibitem[\protect\citeauthoryear{Lowe, Wu, Tamar, Harb, Pieter~Abbeel, and
  Mordatch}{Lowe et~al\mbox{.}}{2017}]%
        {lowe2017multi}
\bibfield{author}{\bibinfo{person}{Ryan Lowe}, \bibinfo{person}{Yi~I Wu},
  \bibinfo{person}{Aviv Tamar}, \bibinfo{person}{Jean Harb},
  \bibinfo{person}{OpenAI Pieter~Abbeel}, {and} \bibinfo{person}{Igor
  Mordatch}.} \bibinfo{year}{2017}\natexlab{}.
\newblock \showarticletitle{Multi-agent actor-critic for mixed
  cooperative-competitive environments}.
\newblock \bibinfo{journal}{\emph{Advances in neural information processing
  systems}}  \bibinfo{volume}{30} (\bibinfo{year}{2017}).
\newblock


\bibitem[\protect\citeauthoryear{Mahajan, Rashid, Samvelyan, and
  Whiteson}{Mahajan et~al\mbox{.}}{2019}]%
        {mahajan2019maven}
\bibfield{author}{\bibinfo{person}{Anuj Mahajan}, \bibinfo{person}{Tabish
  Rashid}, \bibinfo{person}{Mikayel Samvelyan}, {and} \bibinfo{person}{Shimon
  Whiteson}.} \bibinfo{year}{2019}\natexlab{}.
\newblock \showarticletitle{Maven: Multi-agent variational exploration}.
\newblock \bibinfo{journal}{\emph{Advances in Neural Information Processing
  Systems}}  \bibinfo{volume}{32} (\bibinfo{year}{2019}).
\newblock


\bibitem[\protect\citeauthoryear{Rashid, Samvelyan, Schroeder, Farquhar,
  Foerster, and Whiteson}{Rashid et~al\mbox{.}}{2018}]%
        {rashid2018qmix}
\bibfield{author}{\bibinfo{person}{Tabish Rashid}, \bibinfo{person}{Mikayel
  Samvelyan}, \bibinfo{person}{Christian Schroeder}, \bibinfo{person}{Gregory
  Farquhar}, \bibinfo{person}{Jakob Foerster}, {and} \bibinfo{person}{Shimon
  Whiteson}.} \bibinfo{year}{2018}\natexlab{}.
\newblock \showarticletitle{Qmix: Monotonic value function factorisation for
  deep multi-agent reinforcement learning}. In
  \bibinfo{booktitle}{\emph{International conference on machine learning}}.
  PMLR, \bibinfo{pages}{4295--4304}.
\newblock


\bibitem[\protect\citeauthoryear{Samvelyan, Rashid, De~Witt, Farquhar,
  Nardelli, Rudner, Hung, Torr, Foerster, and Whiteson}{Samvelyan
  et~al\mbox{.}}{2019}]%
        {samvelyan2019starcraft}
\bibfield{author}{\bibinfo{person}{Mikayel Samvelyan}, \bibinfo{person}{Tabish
  Rashid}, \bibinfo{person}{Christian~Schroeder De~Witt},
  \bibinfo{person}{Gregory Farquhar}, \bibinfo{person}{Nantas Nardelli},
  \bibinfo{person}{Tim~GJ Rudner}, \bibinfo{person}{Chia-Man Hung},
  \bibinfo{person}{Philip~HS Torr}, \bibinfo{person}{Jakob Foerster}, {and}
  \bibinfo{person}{Shimon Whiteson}.} \bibinfo{year}{2019}\natexlab{}.
\newblock \showarticletitle{The starcraft multi-agent challenge}.
\newblock \bibinfo{journal}{\emph{arXiv preprint arXiv:1902.04043}}
  (\bibinfo{year}{2019}).
\newblock


\bibitem[\protect\citeauthoryear{Sokar, Mocanu, Mocanu, Pechenizkiy, and
  Stone}{Sokar et~al\mbox{.}}{2021}]%
        {sokar2021dynamic}
\bibfield{author}{\bibinfo{person}{Ghada Sokar}, \bibinfo{person}{Elena
  Mocanu}, \bibinfo{person}{Decebal~Constantin Mocanu}, \bibinfo{person}{Mykola
  Pechenizkiy}, {and} \bibinfo{person}{Peter Stone}.}
  \bibinfo{year}{2021}\natexlab{}.
\newblock \showarticletitle{Dynamic sparse training for deep reinforcement
  learning}.
\newblock \bibinfo{journal}{\emph{arXiv preprint arXiv:2106.04217}}
  (\bibinfo{year}{2021}).
\newblock


\bibitem[\protect\citeauthoryear{Su, Chen, Cai, Wu, Gao, Wang, and Lee}{Su
  et~al\mbox{.}}{2020}]%
        {su2020sanity}
\bibfield{author}{\bibinfo{person}{Jingtong Su}, \bibinfo{person}{Yihang Chen},
  \bibinfo{person}{Tianle Cai}, \bibinfo{person}{Tianhao Wu},
  \bibinfo{person}{Ruiqi Gao}, \bibinfo{person}{Liwei Wang}, {and}
  \bibinfo{person}{Jason~D Lee}.} \bibinfo{year}{2020}\natexlab{}.
\newblock \showarticletitle{Sanity-checking pruning methods: Random tickets can
  win the jackpot}.
\newblock \bibinfo{journal}{\emph{Advances in Neural Information Processing
  Systems}}  \bibinfo{volume}{33} (\bibinfo{year}{2020}),
  \bibinfo{pages}{20390--20401}.
\newblock


\bibitem[\protect\citeauthoryear{Sunehag, Lever, Gruslys, Czarnecki, Zambaldi,
  Jaderberg, Lanctot, Sonnerat, Leibo, Tuyls, et~al\mbox{.}}{Sunehag
  et~al\mbox{.}}{2018}]%
        {sunehag2018value}
\bibfield{author}{\bibinfo{person}{Peter Sunehag}, \bibinfo{person}{Guy Lever},
  \bibinfo{person}{Audrunas Gruslys}, \bibinfo{person}{Wojciech~Marian
  Czarnecki}, \bibinfo{person}{Vin{\'\i}cius~Flores Zambaldi},
  \bibinfo{person}{Max Jaderberg}, \bibinfo{person}{Marc Lanctot},
  \bibinfo{person}{Nicolas Sonnerat}, \bibinfo{person}{Joel~Z Leibo},
  \bibinfo{person}{Karl Tuyls}, {et~al\mbox{.}}}
  \bibinfo{year}{2018}\natexlab{}.
\newblock \showarticletitle{Value-Decomposition Networks For Cooperative
  Multi-Agent Learning Based On Team Reward}. In
  \bibinfo{booktitle}{\emph{AAMAS}}.
\newblock


\bibitem[\protect\citeauthoryear{Tan}{Tan}{1993}]%
        {tan1993multi}
\bibfield{author}{\bibinfo{person}{Ming Tan}.} \bibinfo{year}{1993}\natexlab{}.
\newblock \showarticletitle{Multi-agent reinforcement learning: Independent vs.
  cooperative agents}. In \bibinfo{booktitle}{\emph{Proceedings of the tenth
  international conference on machine learning}}. \bibinfo{pages}{330--337}.
\newblock


\bibitem[\protect\citeauthoryear{Terry, Grammel, Hari, and Santos}{Terry
  et~al\mbox{.}}{2020}]%
        {terry2020parameter}
\bibfield{author}{\bibinfo{person}{Justin~K Terry}, \bibinfo{person}{Nathaniel
  Grammel}, \bibinfo{person}{Ananth Hari}, {and} \bibinfo{person}{Luis
  Santos}.} \bibinfo{year}{2020}\natexlab{}.
\newblock \showarticletitle{Parameter sharing is surprisingly useful for
  multi-agent deep reinforcement learning}.
\newblock  (\bibinfo{year}{2020}).
\newblock


\bibitem[\protect\citeauthoryear{Wang, Qin, Bai, Zhang, and Fu}{Wang
  et~al\mbox{.}}{2021}]%
        {wang2021recent}
\bibfield{author}{\bibinfo{person}{Huan Wang}, \bibinfo{person}{Can Qin},
  \bibinfo{person}{Yue Bai}, \bibinfo{person}{Yulun Zhang}, {and}
  \bibinfo{person}{Yun Fu}.} \bibinfo{year}{2021}\natexlab{}.
\newblock \showarticletitle{Recent Advances on Neural Network Pruning at
  Initialization}.
\newblock \bibinfo{journal}{\emph{arXiv e-prints}} (\bibinfo{year}{2021}),
  \bibinfo{pages}{arXiv--2103}.
\newblock


\bibitem[\protect\citeauthoryear{Yang, Luo, Li, Zhou, Zhang, and Wang}{Yang
  et~al\mbox{.}}{2018}]%
        {yang2018mean}
\bibfield{author}{\bibinfo{person}{Yaodong Yang}, \bibinfo{person}{Rui Luo},
  \bibinfo{person}{Minne Li}, \bibinfo{person}{Ming Zhou},
  \bibinfo{person}{Weinan Zhang}, {and} \bibinfo{person}{Jun Wang}.}
  \bibinfo{year}{2018}\natexlab{}.
\newblock \showarticletitle{Mean field multi-agent reinforcement learning}. In
  \bibinfo{booktitle}{\emph{International conference on machine learning}}.
  PMLR, \bibinfo{pages}{5571--5580}.
\newblock


\end{thebibliography}

%%%%%%%%%%%%%%%%%%%%%%%%%%%%%%%%%%%%%%%%%%%%%%%%%%%%%%%%%%%%%%%%%%%%%%%%

\onecolumn

%%%%%%%%%%%%%%%%%%%%%%%%%%%%%%%%%%%%%%%%%%%%%%%%%%%%%%%%%%%%%%%%%%%%%%%%%%%%%%%%%%%%%%%%%%%
\section*{Appendix A: Experimental Results}
\label{sec:appendix A }

\begin{figure*}[h]
\begin{center}
\begin{tabular}{ccc}
     % uncomment the next lines, and give the right ps files
     %\includegraphics[width=0.23\textwidth]{figures/MW_N2.png} &
     \includegraphics[width=0.3\textwidth]{figures/5m_vs_6m.pdf} & \hspace{-3ex}
     \includegraphics[width=0.31\textwidth]{figures/results_smac_8m_vs_9m.pdf} &
     \hspace{-3ex}
     \includegraphics[width=0.28\textwidth]{figures/results_smac_mmm2.pdf} \\
     (a) 5m vs 6m (SMAC) &\hspace{-3ex} (b) 8m vs 9m (SMAC) &\hspace{-3ex} (c) MMM2 (SMAC) \\
     \includegraphics[width=0.29\textwidth]{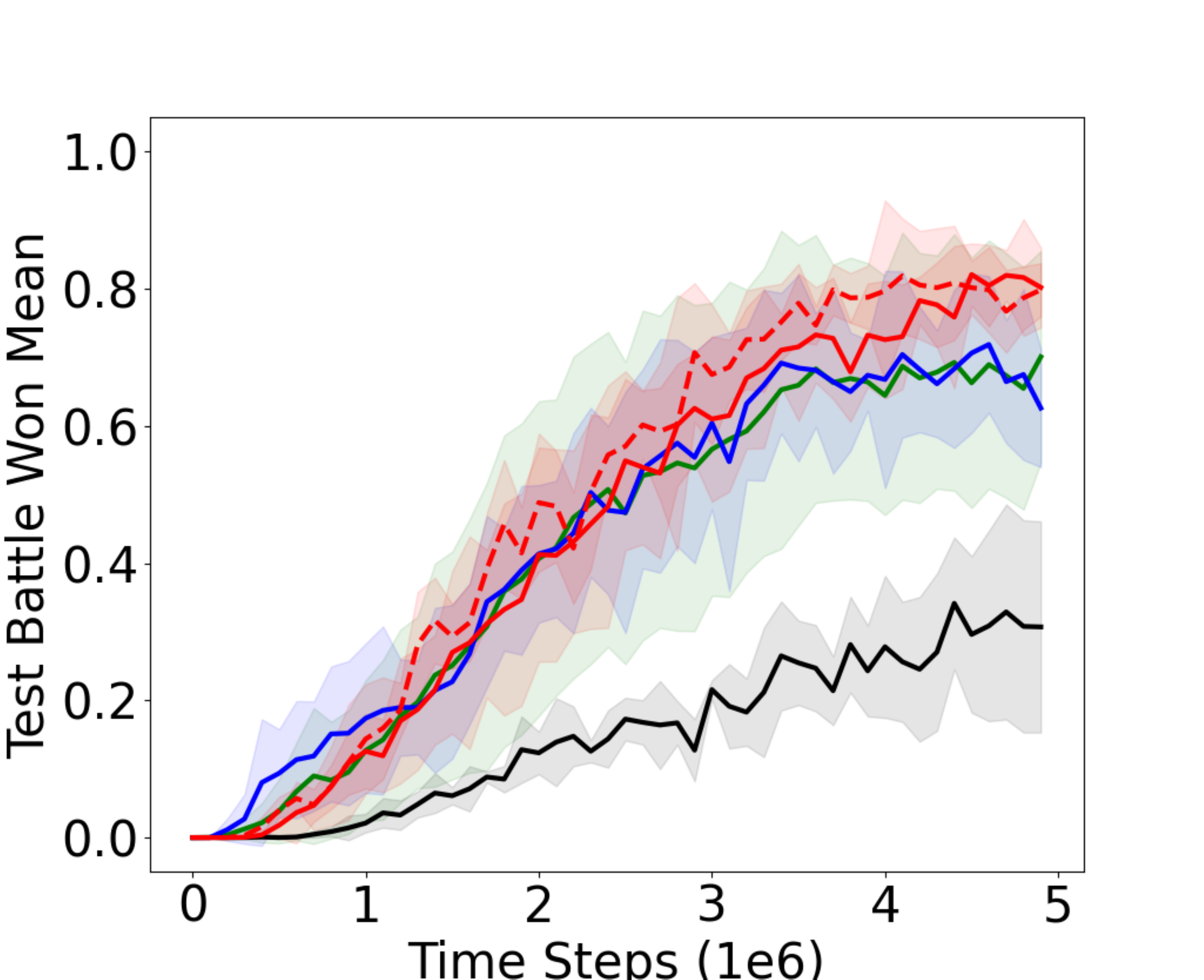} &
     \hspace{-3ex}
    \includegraphics[width=0.295\textwidth]{figures/results_lbf_n6v6.pdf} &
     \hspace{-3ex}
    \includegraphics[width=0.3\textwidth]{figures/result_lbf_n6v7.pdf}\\
    (d) 27m vs 30m (SMAC) &\hspace{-3ex} (f) LBF1 &\hspace{-3ex} (f) LBF2  \\
\end{tabular}
\caption{Performance of FuPS (green), FuPS+id (blue), SePS (black), SNP-PS (red, ours), and SNP-PS+id (red-dotted, ours) on SMAC environments and LBF environments. More results are provided in Appendix.}
\vspace{-1em}
\label{fig:results}
\end{center}
\end{figure*}

\end{document}